\documentclass[aps,pra,twocolumn,superscriptaddress,showpacs]{revtex4-2}

\usepackage[utf8]{inputenc}

\usepackage{array}
\usepackage{amsmath}
\usepackage{graphicx}
\usepackage{epstopdf}
\usepackage{color}
\usepackage{amsfonts}
\usepackage{subfigure}
\usepackage{amssymb}
\usepackage{graphicx}
\usepackage{makecell}
\usepackage[colorlinks,citecolor=blue]{hyperref}
\usepackage{changes}
\newcommand{\onecolm}{
  \end{multicols}
  \vspace{-3.5ex}
  \noindent\rule{0.5\textwidth}{0.1ex}\rule{0.1ex}{2ex}\hfill
}
\newcommand{\twocolm}{
  \hfill\raisebox{-1.9ex}{\rule{0.1ex}{2ex}}\rule{0.5\textwidth}{0.1ex}
  \vspace{-4ex}
  \begin{multicols}{2}
}

\begin{document}

\title{The Intrinsic Connection between Dynamical Phase Transitions and Magnetization in the 1D XY Model}

\date{\today}

\author{Lin-Yue Luo}
\affiliation {Key Laboratory of Atomic and Subatomic Structure and Quantum Control (Ministry of Education), Guangdong Basic Research Center of Excellence for Structure and Fundamental Interactions of Matter, School of Physics, South China Normal University, Guangzhou 510006, China}
\affiliation {Guangdong Provincial Key Laboratory of Quantum Engineering and Quantum Materials, Guangdong-Hong Kong Joint Laboratory of Quantum Matter, Frontier Research Institute for Physics, South China Normal University, Guangzhou 510006, China}

\author{Wei-Lin Li}
\affiliation {Key Laboratory of Atomic and Subatomic Structure and Quantum Control (Ministry of Education), Guangdong Basic Research Center of Excellence for Structure and Fundamental Interactions of Matter, School of Physics, South China Normal University, Guangzhou 510006, China}
\affiliation {Guangdong Provincial Key Laboratory of Quantum Engineering and Quantum Materials, Guangdong-Hong Kong Joint Laboratory of Quantum Matter, Frontier Research Institute for Physics, South China Normal University, Guangzhou 510006, China}

\author{Bao-Ming Xu}
\email{corresponding author: xubm2018@163.com}
\affiliation{Institute of Biophysics, Dezhou University, Dezhou 253023, China}

\author{Zhi Li}
\email{corresponding author: lizphys@m.scnu.edu.cn}
\affiliation {Key Laboratory of Atomic and Subatomic Structure and Quantum Control (Ministry of Education), Guangdong Basic Research Center of Excellence for Structure and Fundamental Interactions of Matter, School of Physics, South China Normal University, Guangzhou 510006, China}
\affiliation {Guangdong Provincial Key Laboratory of Quantum Engineering and Quantum Materials, Guangdong-Hong Kong Joint Laboratory of Quantum Matter, Frontier Research Institute for Physics, South China Normal University, Guangzhou 510006, China}

\begin{abstract}
In this manuscript, we study the quench dynamics of a transverse-field XY model starting from coherent Gibbs states. The results reveal that the initial strength of magnetization plays a crucial role in the emergence of dynamical quantum phase transitions. In concrete terms, when quenching within the same phase, through the properties of observables such as Fisher zeros and magnetization, we show that the stronger the initial magnetization, the more difficult the emergence of dynamical quantum phase transitions. The underlying mechanism is that the strong initial magnetization provides a directional effect, which inhibits the spin flipping in the process of quantum quench, making the dynamical quantum phase transition difficult to emerge. Since dynamical quantum phase transitions can be experimentally realized in various artificial systems, we hope that the physics predicted here can be experimentally verified in tabletop platforms.
\end{abstract}
%\pacs{03.65.Ud, 05.30.Rt, 03.67.-a, 42.50.-p}

\maketitle

%----------------------------------------------------------------------------------------------------

\section{INTRODUCTION}
The rapid advances in quantum simulation platforms have greatly propelled the study of non-equilibrium dynamics. Within this field, dynamical quantum phase transitions (DQPTs) have attracted considerable interest~\cite{PRL.2013.heyl,eisert2015quantum,RevModPhys.2011.Polkovnikov,PRL.2005.Zurek,heyl2018dynamical,heyl2019dynamical,marino2022dynamical,zvyagin2016dynamical}. DQPTs can emerge in the process of quench dynamics. Previous studies show that DQPT can be classified into two types~\cite{PRL.2018.Zunkovic,PRL.2023.Corps}. The first type of DQPT, denoted as DQPT-OP, is characterized by a discontinuity in the long-time limit value of an order parameter (or certain correlation functions) following a quench, serving as a dynamical hallmark of the underlying phase transition~\cite{eckstein2008nonthermal,eckstein2009thermalization,moeckel2008interaction,moeckel2010crossover,Sciolla2010Infinite-Dimensional,Sciolla_2013,gambassi2011quantum,Maraga2015aging,chandran2013equilibration,smacchia2015exploring,halimeh2017prethermalization,mori2018thermalization,muniz2020exploring,smale2019observation,tian2020observation}. The second type of DQPT, referred to as DQPT-LE, is characterized by non-analyticities in the time evolution of the rate function of the Loschmidt echo~\cite{arxiv.2024.cao,NJP.2023.Cheraghi,PRB.2016.Budich,PRB.2017.Bhattacharya,PRB.2017.heyl,PRB.2020.cao,PRL.2013.heyl,PRL.2014.heyl,PRL.2015.hely,ScientificReports.2019.Jafari,PRB.2014.Hickey}. Theoretically, previous works show that DQPT-LE can emerge in transverse-field Ising model~\cite{PRL.2013.heyl,PRB.2017.Halimeh,PRL.2018.LANG}, XY model~\cite{PRB.2014.Vajna,ScientificReports.2020.Porta,CPB.2022.cao,PRE.2016.Divakaran,PRB.2025.ZENG}, Kitaev honeycomb models~\cite{PRB.2015.Schmitt}, Floquet systems~\cite{EPL.2014.Sharma,PRB.2019.YANGK,PRB.2020.Zamani,JOP.2021.zhoulw,PRA.2021.JAFARI,NatureC.2021.Hamazaki,PRB.2022.Zamani,PhysicaA.2022.Luan,PRB.2022.jafari}, and topological systems~\cite{PRB.2015.Vajna,PRL.2017.Jafari,PRB.2018.Sedlmayr,PRL.2019.Zache,PRR.2021.Okugawa,PRB.2019.Jafari,PRB.2020.Masowski}, etc~\cite{PRB.2015.Sharma,PRE.2017.Zauner,PRB.2017.Homrighausen,PRB.2017.Obuchi,PRB.2017.Dutta,PRR.2020.Halimeh,PRA.2018.zhoulw,NJP.2021.zhoulongwen,PRB.2022.Mondal,PRB.2023.Mondal,PRB.2018.Lang,PRB.2020.Hou,PRA.2020.Kyaw,PRB.2018.Mera,PRB.2018.SedlmayrFate,PRL.2020.Link,PRA.2021.Hou,PRB.2017.Weidinger,JPA.2016.Jafari,PRB.2019.Abdi}. Experimentally, DQPT-LE has been realized on various quantum simulators, such as ultracold atoms~\cite{NaturePhysics.2018.flaschner}, trapped ions~\cite{PRL.2017.Jurcevic}, superconducting quantum circuits~\cite{PRAp.2019.GUOXY}, etc~\cite{APL.2020.Chen,PRL.2019.Wang,Light.2020.Xu,LSC.2025.Zhang}.

On the other hand, the underlying mechanism of the emergence of DQPT-LE remains an open question. Initially, it was believed that DQPT-LE originated from equilibrium quantum criticality. This was because DQPT-LE was first discovered in the study of the quench dynamics \cite{2016Essler,Annual.2018.Mitra} across the phase transition points of different equilibrium states~\cite{PRL.2013.heyl,PRB.2013.Karrasch,PRB.2014.Kriel,PRB.2018.Bhattacharjee,PRB.2019.Zhou}. However, recent research has shown that quenching within the same phase can also induce DQPT-LE~\cite{PRB.2014.Andraschko,PRB.2014.Vajna,PRB.2015.Vajna,PRB.2020.Uhrich,PRB.2022.Rossi,SciPostPhys.2021.Liska,PRB.2024.Wong}. In other words, DQPT-LE is a truly non-equilibrium phenomenon, and it is fundamentally unrelated to equilibrium critical points. While their topological features have been identified~\cite{PRB.2016.Sharma,PRA.2018.Qiu,PRB.2019.Mendl,PRB.2021.Sadrzadeh,PRB.2021.Yu,PRA.2022.Naji,PRL.2016.Huang}, a comprehensive understanding of their dynamical principles is still not uncovered.
Furthermore, the magnetization dynamics have long served as a fundamental probe for DQPTs. In the context of DQPT-OP, the relaxation of magnetization (or its related two-point correlation functions) often follows a power-law decay ($t^{-\alpha}$), with the exponent $\alpha$ being primarily determined by the phase of the post-quench Hamiltonian~\cite{SCIPOST.2016.Tatjana,PRB.2022.Makki,PRA.2024.CAO,PRB.2024.CAO2}. Intriguingly, in the pioneering work of DQPT-LE, the period of the oscillations of the magnetization was found to coincide with the period of the critical times of DQPT-LE within numerical accuracy~\cite{PRL.2013.heyl}.

So far, most mainstream research has focused on the properties of DQPTs encoded in the post-quench evolution of magnetization~\cite{PRB.2019.Lerose,PRR.2022.Hashizume,zhang2017observation,PRB.2018.Lang.qcp}, while the influence of the initial magnetization on DQPTs has remained largely unexplored~\cite{halimeh2017prethermalization}. As a key quantity that reflects the collective spin order of the system, the initial magnetization is crucial for DQPTs. This motivates our study of its influence on the quench dynamics. This work studies the quench dynamics of a prototypical spin chain: The one-dimensional anisotropic XY model. The initial state is prepared in the coherent Gibbs state~\cite{PRA.2018.XU,CTP.2024.XU}. This approach allows us to investigate how the initial magnetization affects the emergence of DQPT-LE. In the following discussions, `DQPT' will refer specifically to DQPT-LE for convenience.

The rest of this manuscript is organized as follows: Sec.~\ref{sec:model} introduces the model and the corresponding phase diagram. In Sec.~\ref{sec:DQPT}, we discuss the influence of coherent Gibbs states on the DQPT. In Sec.~\ref{sec:correlation}, we discuss the relationship between the strength of magnetization and the emergence of DQPT. Sec.~\ref{sec:conclusion} is the conclusion.
%----------------------------------------------------------------------------------------------------

\section{MODEL AND the quench path on the global PHASE DIAGRAM}
\label{sec:model}
We start with the transverse-field XY model. The corresponding Hamiltonian reads~\cite{PRA.2024.CAO}
%%%
%%%
\begin{equation}
\hat{H}=-\frac{J}{2}\sum_{j=1}^{N}\left( \frac{1+\gamma}{2} \sigma_{j}^{x} \sigma_{j+1}^{x}+\frac{1-\gamma}{2} \sigma_{j}^{y} \sigma_{j+1}^{y}+\lambda \sigma_{j}^{z}\right),
\label{Eq1}
\end{equation}
%%%
%%%
where $J$ denotes the longitudinal spin-spin coupling strength. $\gamma \in [-1, 1]$ is the anisotropic parameter, which controls the coupling along the $x$ and $y$ directions. Without loss of generality, we set $J=1$ as the energy unit. $\sigma^{\alpha}_{j}(\alpha=x,y,z)$ is the spin-1/2 Pauli matrix defined on the lattice site $j$. The periodic boundary condition is employed, i.e., $\hat{\sigma}_{N+1}=\hat{\sigma}_{1}$.

%%%%%%
%%%%%%
\begin{figure}[tbhp] \centering
\includegraphics[width=6.5cm]{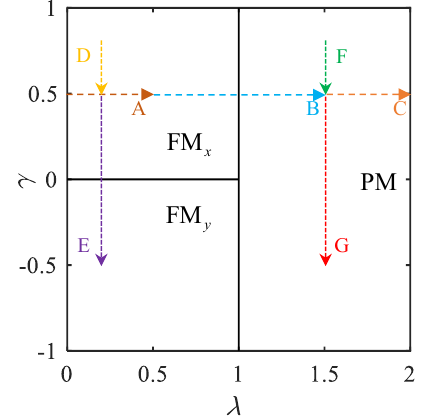}
\caption{Phase diagram of the transverse-field XY model. Solid lines are the critical lines, which separate three gapped phases. Dashed arrows denote the different quench paths A-G.}
\label{figphasegram}
\end{figure}
%%%%%%
%%%%%%

The global phase diagram is shown in Fig.~\ref{figphasegram}. The previous study shows that the excitation spectrum is gapless at $\lambda=1$ or at $\gamma=0$~\cite{PRA.2024.CAO,PRB.2014.Vajna}. As shown in Fig.~\ref{figphasegram}, the Ising transition, i.e., the phase transition from the ferromagnetic ($\mathrm{FM}$) phase to the paramagnetic ($\mathrm{PM}$) phase, can occur by tuning the external magnetic field $\lambda$. The corresponding quantum critical point $\lambda_c=1$. The anisotropic phase transition, i.e., the phase transition between $\mathrm{FM}_x$ phase and $\mathrm{FM}_{y}$ phase, can be achieved by manipulating the parameter $\gamma$. The corresponding critical point $\gamma_c=0$.
In this manuscript, we explore dynamical behaviors under a variety of quench protocols, encompassing both intra-phase quenches and quenches across critical lines. The corresponding quench paths are marked in the phase diagram (see Fig.~\ref{figphasegram}). Note that the Hamiltonian~\eqref{Eq1} is integrable. Then, one can map it to a free fermion model, in which we can obtain the corresponding exact solution. By applying the Jordan-Wigner transformation and the Fourier transformation, one can get
%%%
%%%
\begin{equation}
\hat{H}=\sum_{k>0}\left(\begin{array}{ll}\hat{c}_{k}^{\dagger} & \hat{c}_{-k}\end{array}\right)\hat{H}_{k}\binom{\hat{c}_{k}}{\hat{c}_{-k}^{\dagger}},
\label{Eq2}
\end{equation}
%%%
%%%
with
%%%
%%%
\begin{equation}
\hat{H}_{k}=\left(\begin{array}{cc}-\lambda-\cos k & i \gamma \sin k \\ -i \gamma \sin k & \lambda+\cos k\end{array}\right),
\label{EqHk}
\end{equation}
%%%
%%%
where $\hat{c}_k$ ($\hat{c}_k^{\dagger}$) denotes the fermion annihilation (creation) operator for mode $k=(2n-1)\pi/N$ with $n=1,2 \cdots N/2$. $\hat{H}_k$ acts on a two-dimensional Hilbert space generated by $\left\{\lvert 1 \rangle=\hat{c}_k^\dagger \hat{c}_{-k}^\dagger \lvert 0 \rangle, \lvert 0 \rangle \right\}$, where $|0\rangle$ is the fermion vacuum state of the Jordan-Wigner fermions $\hat{c}_{k}$ and $\hat{c}_{-k}$, and can be represented in that basis by a $2\times2$ matrix. Then, one can get the instantaneous eigenvalues, i.e., $\varepsilon_k^{\pm}=\pm\varepsilon_k$ with
%%%
%%%
\begin{equation}
    \varepsilon_k=\sqrt{(\lambda+\cos k)^2+\gamma^2\sin^2 k}.
    \label{EQeigenvalues}
\end{equation}
%%%
%%%
The corresponding eigenvectors are
%%%
%%%
\begin{equation}
\begin{aligned}
\left|\varepsilon_k^+\right\rangle &= -i\sin\theta_k\, \lvert 1 \rangle + \cos\theta_k\, \lvert 0 \rangle, \\
\left|\varepsilon_k^-\right\rangle &= \cos\theta_k\, \lvert 1 \rangle - i\sin\theta_k\, \lvert 0 \rangle,
\end{aligned}
\end{equation}
%%%
%%%
where the $\theta_k$ is defined by
%%%
%%%
\begin{equation}
e^{i\theta_k} = \frac{-\lambda-\cos k-\varepsilon_k+i \gamma \sin k }{\sqrt{\gamma^2\sin^2 k+ (-\lambda-\cos k-\varepsilon_k)^2}}.
\label{EQphase}
\end{equation}
%%%
%%

The system is initially prepared in state $|\psi(0)\rangle$ for time $t < 0$. At $t=0$, the system undergoes a sudden quench, with the Hamiltonian switching from $\hat{H}(\lambda,\gamma)$ to $\hat{H}(\lambda',\gamma')$. The subsequent dynamics is governed by the operator $\hat{U}(t)=e^{-i\hat{H}(\lambda',\gamma')t}$ ($\hbar = 1$). One can obtain the time-evolved state $|\psi(t)\rangle$ as
%%%
%%%
\begin{equation}\label{}
  |\psi(t)\rangle=e^{-i\hat{H}(\lambda',\gamma')t}|\psi(0)\rangle.
\end{equation}
%%%
%%%

Following the quench, the Loschmidt amplitude is given by
%%%
%%%
\begin{equation}
\mathcal{G}\left(t\right)=\langle \psi(0) | \psi(t) \rangle=\langle \psi(0)| e^{-i\hat{H}(\lambda',\gamma')t} | \psi(0) \rangle.
\label{LA}
\end{equation}
%%%
%%%
The corresponding Loschmidt echo, which quantifies the probability of the time-evolved state returning to the initial state, is given by $\mathcal{L}(t)=|\mathcal{G}(t)|^2$. A DQPT is signaled by a vanishing Loschmidt echo $\mathcal{L}(t)=0$, which occurs when the time-evolved state becomes orthogonal to the initial state. In the XY spin chain, this orthogonality fundamentally indicates that a collective spin flip has occurred during the quench dynamics~\cite{ScientificReports.2018.Bandyopadhyay}.
Eq.~\ref{LA} shows that this dynamical process is governed by the initial state and the post-quench Hamiltonian. Consequently, for a fixed quench protocol, the initial spin polarization, quantified by the magnetization, plays the decisive role. It directly determines the ease of spin flipping during the quench and thereby controls the emergence of a DQPT. Hence, this work will delve into the relationship between the initial magnetization and DQPTs.

To this end, we consider the coherent Gibbs state as the initial state, which can be expressed by
%%%
%%%
\begin{equation}
\left|\psi(0)\right\rangle = \bigotimes_k \lvert \psi_k(0) \rangle,
\end{equation}
%%%
%%%
where
%%%
%%%
\begin{equation}
\label{Eq initial state}
\left|\psi_k(0)\right\rangle = \sqrt{\frac{e^{-\beta \varepsilon_k}}{Z_k}} \left| \varepsilon_k^+\right\rangle + \sqrt{\frac{e^{\beta \varepsilon_k}}{Z_k}} e^{i\phi_k} \left| \varepsilon_k^- \right\rangle,
\end{equation}
%%%
%%%
with $Z_k= 2\cosh[\beta \varepsilon_k]$ referring to the partition function. In the expression above, $\beta=1/(k_BT)$. Without loss of generality, we set the Boltzmann constant $k_B=1$ in the following calculation.
Note that the density matrices of the coherent Gibbs state and the traditional Gibbs state share identical diagonal elements in the energy eigenbasis, leading to indistinguishable energy distributions. The key distinction lies in the off-diagonal elements, where the coherent Gibbs state introduces quantum coherence. The energy distribution of the traditional Gibbs state is governed by the actual inverse temperature, whereas in our coherent Gibbs state, it is governed by the parameter $\beta$. Since both states exhibit identical energy distributions for a given $\beta$, this parameter plays a role analogous to the inverse temperature in a traditional Gibbs ensemble. We therefore refer to $\beta$ as the ``effective inverse temperature’’.
The parameter $\phi_k$ denotes the relative phase associated with momentum mode $k$. For simplicity, we assume a uniform relative phase across all modes at the same energy level, i.e., $\phi_k\equiv\phi$.

\section{The effects of $\beta$ on DQPTs}\label{sec:DQPT}
In this section, we examine how the emergence of the DQPTs is influenced by the initial state, which we control by tuning $\beta$. We investigate this dependence under two classes of quench paths defined in Fig.~\ref{figphasegram}: Transverse field quenches and anisotropy parameter quenches.

Classical equilibrium phase transitions are characterized by the zeros of the partition function (e.g. Lee-Yang zeros), which signal non-analyticities in thermodynamic observables. This framework can be extended to non-equilibrium dynamics, providing a direct way to diagnose DQPTs through Fisher zeros. Therefore, we adopt this approach to identify the occurrence of DQPTs in our system. The Fisher zeros of the Loschmidt amplitude $\mathcal{G}(t)$, analytically continued to the complex $z$ plane (i.e., by replacing $it$ in Eq.~\ref{LA} with $z$), satisfy the expression
%%%
%%%
\begin{equation}
\mathcal{G}(z)=\left\langle\psi\left(0\right)\right|e^{-z\hat{H}(\lambda',\gamma')}\left|\psi\left(0\right)\right\rangle=0,
\label{Zz}
\end{equation}
%%%
%%%
where $z\in \mathbb{C}$. Through a straightforward calculation, we obtain the analytical expression of the Fisher zeros~\cite{CTP.2024.XU,fisher1965boulder}, i.e.,
%%%
%%%
\begin{widetext}
\begin{equation}
\begin{aligned}
z_n(k) = \frac{1}{2\varepsilon_k(\lambda',\gamma')} \bigg[ \ln\bigg|
\frac{e^{-\beta\varepsilon_k(\lambda,\gamma)} \cos^2(\Delta_{\theta_k})
+ e^{\beta\varepsilon_k(\lambda,\gamma)} \sin^2(\Delta_{\theta_k})
+ \sin(2\Delta_{\theta_k})\sin\phi}
{e^{-\beta\varepsilon_k(\lambda,\gamma)} \sin^2(\Delta_{\theta_k})
+ e^{\beta\varepsilon_k(\lambda,\gamma)} \cos^2(\Delta_{\theta_k})
- \sin(2\Delta_{\theta_k})\sin\phi}
\bigg| + i(2n+1)\pi \bigg],
\label{z_n}
\end{aligned}
\end{equation}
\end{widetext}
%%%
%%%
where $\Delta_{\theta_k}=\theta_k-\theta_k'$.
In a finite system, the Fisher zeros appear as discrete points in the complex $z$ plane. With increasing system size, the density of Fisher zeros increases. In the thermodynamic limit, the Fisher zeros coalesce into continuous lines indexed by an integer $n \in \mathbb{Z}$ in the complex $z$ plane. In the following analysis, we focus on the $n=0$ branch.
A DQPT occurs when the line of Fisher zeros intersects the imaginary axis in the complex $z$ plane. The intersection defines the critical times of the DQPTs, where the time-evolved state becomes orthogonal to the initial state, which coincides with a collective spin flip. Moreover, each point at which a Fisher-zero line crosses the imaginary axis is associated with a critical momentum $k^*$. Each $k^*$ corresponds to a distinct type of DQPT in the quench process, specifically characterized by a unique periodicity in critical time.
%which corresponds to the first critical time and is typically the most relevant for observing DQPTs.

\subsection{Quench of the transverse field $\lambda$}
\begin{figure}[tphp] \centering%[tph]
\includegraphics[width=7.5cm]{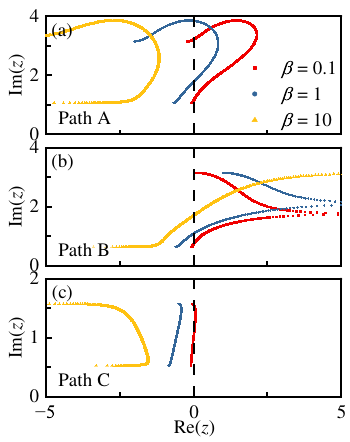}
\caption{Fisher zeros for quenches of the transverse field along quench paths A ($\lambda=0\rightarrow0.5$), B ($\lambda=0.5\rightarrow1.5$), and C ($\lambda=1.5\rightarrow2.0$) with different $\beta$. Throughout, $\phi=-\pi/2$ and $\gamma=0.5$.}
\label{FIG2lamquench}
\end{figure}

We first calculate the Fisher zeros for quenching paths along the transverse field $\lambda$, i.e., paths A, B, and C in Fig.~\ref{figphasegram}. Since there is no qualitative difference for the case of $\mathrm{FM}_y$, we will only present the results for the case of $\mathrm{FM}_x$~(see Fig.~\ref{FIG2lamquench}).

Firstly, for a given quench protocol within the $\mathrm{FM}_x$ phase (path A), the lines of Fisher zeros cross the imaginary axis for the case of $\beta =0.1,1$, whereas in the case of $\beta=10$, the corresponding line of Fisher zeros is always unable to cross the imaginary axis ~[see Fig.~\ref{FIG2lamquench}(a)]. From Eq.~\ref{Eq initial state}, reducing $\beta$ is shown to enhance quantum coherence in the coherent Gibbs state. These results demonstrate that for a fixed quench protocol, the resulting enhanced coherence opens a pathway to DQPTs that are forbidden when quenching from the ground state.

%%%%%%
%%%%%%
%%%%%%
%%%%%%

Secondly, for a quench path that crosses a critical point, we observe that the line of Fisher zeros always crosses the imaginary axis [see Fig.~\ref{FIG2lamquench}(b)]. This finding is consistent with previous studies~\cite{PRB.2014.Vajna,PRL.2013.heyl,PRB.2013.Karrasch}, which have shown that quenches connected to equilibrium critical points tend to induce DQPTs.

Thirdly, one can observe that the DQPTs also emerge for the quench path within the $\mathrm{PM}$ region for $\beta=0.1$ [see Fig.~\ref{FIG2lamquench}(c)].

Furthermore, one can find that during the same-phase quench [Fig.~\ref{FIG2lamquench}(a) and (c)], the line of Fisher zeros monotonously approaches the imaginary axis with a decreasing $\beta$. This implies that for a given quench protocol, there exists a critical $\beta$ at which the line of Fisher zeros just touches the imaginary axis. This contact point signals the emergence of a DQPT.

\subsection{Quench of anisotropy $\gamma$}
\begin{figure}[tphp] \centering
\includegraphics[width=7.5cm]{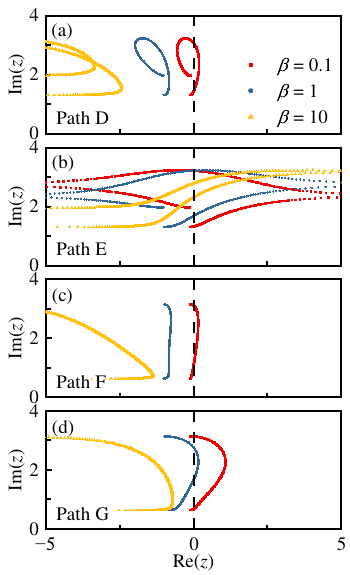}
\caption{Fisher zeros for anisotropy quench paths. The path D (E) satisfies $\gamma=0.8\rightarrow0.5$ ($\gamma=0.5\rightarrow-0.5$) with fixed $\lambda=0.2$. Path F (G) follows the identical $\gamma$ quench as path D (E), but at a different $\lambda$=1.5. Throughout, $\phi=-\pi/2$.}
\label{FIG3gamquench}
\end{figure}

Having discussed transverse field quenches, we now turn to the DQPTs induced by quenching the anisotropy parameter $\gamma$. We focus on quench paths D, E, F, G of Fig.~\ref{figphasegram}. The results reveal that the lines of Fisher zeros corresponding to all the quenching paths will cross the imaginary axis, which means DQPT can be induced by $\beta$ in all cases~(see Fig.~\ref{FIG3gamquench}). It can be seen clearly that the results of quenching the anisotropic parameter are similar to those obtained from quenching the transverse field parameter before. In concrete terms, controlling $\beta$ can induce DQPT in situations where it was originally impossible for the emergence of DQPT. For cases that cross the phase transition point, the same reason, i.e., the promoting effect of $\beta$, ensures that the emergence of DQPT is not disrupted.

%%%%%%
%%%%%%
%%%%%%
%%%%%%

%%%%%%%%%%%%%%%%%%%%%%%%%%%%%%%5修改至此 20251027
Based on the results of the two typical quench paths mentioned above, one can find that the coherent Gibbs state brings new degrees of freedom to the quenching dynamics, i.e., a tunable $\beta$. A reduced $\beta$ enables the system to exhibit DQPTs that are previously unobservable under the same intra-phase quench protocol when starting from the traditional ground state. Carefully check expressions Eq.~\eqref{EQeigenvalues}-\eqref{Eq initial state}, one can find that essentially, $\beta$ governs the quantum coherence in the coherent Gibbs state by tuning the energy-level populations. In the limit of $\beta \rightarrow \infty$, the state reduces to the traditional ground state, devoid of quantum coherence. Conversely, as $\beta \rightarrow 0$, the populations of various energy levels become equal, maximizing the quantum coherence for an appropriate relative phase $\phi$.
Note that, as for the case of the same-phase quench, one can find that the decrease in $\beta$ enables us to discover a new $\beta$-dependent DQPT to emerge. The next section examines how $\beta$ affects the magnetization, providing insight into the origin of this novel DQPT.

%Since DQPTs arise from spin flips and $\beta$ only affects the initial state, this directs our focus to the role of $\beta$ in setting the initial magnetization. This magnetization is a key factor governing the subsequent dynamics.

\section{The effect of Magnetization on DQPT}\label{sec:correlation}
\begin{figure}[tph] \centering%[tph]
\includegraphics[width=7cm]{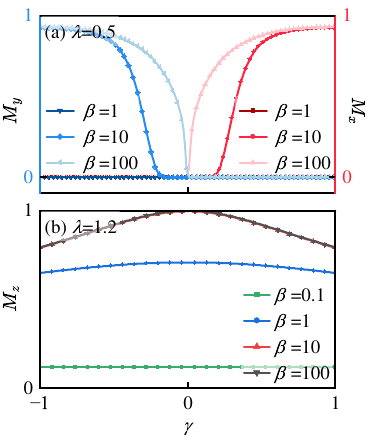}
\caption{Magnetization versus $\gamma$ for (a) $M_x$, $M_y$ with $\lambda=0.5$, and (b) $M_z$ with $\lambda=1.2$, respectively. Throughout, $\phi=-\pi/2$, and the values of $\beta$ are marked.}
\label{FIG4temperatureM}
\end{figure}

%%%
%%%
The magnetization along the $\alpha$-direction ($\alpha = x, y, z$) is defined as
\begin{equation}
    M_\alpha=\langle\sigma^\alpha_j\rangle.
\end{equation}
%%%
%%%
However, due to the $Z_2$ symmetry of the system~\cite{PFEUTY197079,barouchStatisticalMechanicsModel1971,Lieb2004}, the order parameters $M_x$ and $M_y$ cannot be directly obtained (see Appendix~\ref{APDIX:Magnetization} for details).
For the FM phase, under the condition of $\gamma>0$ ($\gamma<0$), the polarization of the system is along the $x$ ($y$) direction. For the PM phase, the corresponding polarization is determined by the transverse field, i.e., along the $z$-direction. We calculate the corresponding strength of magnetization $M_x$, $M_y$ and $M_z$ versus $\gamma$ under different $\beta$ (see Fig.~\ref{FIG4temperatureM}).

%%%%%%
%%%%%%
%%%%%%
%%%%%%

In the limit $\beta\rightarrow\infty$, the system reduces to the ground state (see the $\beta=100$ lines in Fig.~\ref{FIG4temperatureM}), which agrees well with previous results~\cite{JPA.2010.Zhong}. This corresponds to the case of max magnetization, i.e., $M_x\rightarrow 1$ or $M_y\rightarrow 1$. As shown in the Fig.~\ref{FIG4temperatureM},  the magnetizations $M_x$, $M_y$, and $M_z$ will decrease and eventually tend towards zero with a decreasing $\beta$. This phenomenon demonstrates that reducing $\beta$ suppresses the magnetization, which enhances the susceptibility of the system to collective spin flips during a quench, thus promoting DQPTs. By comparing the trends of magnetization as a function of $\gamma$ for different fixed values of $\beta$ in Fig.~\ref{FIG4temperatureM} (a) and (b), we find that under the condition of $\beta=1$, the magnetization of FM phase can be completely suppressed ($M_x\rightarrow 0$ or $M_y\rightarrow 0$), while in the PM phase, the magnetization can only be strongly suppressed under the condition of $\beta=0.1$. In other words, compared to the FM phase, a smaller value of $\beta$ is required to completely destroy the magnetization in the PM phase.
This indicates that the magnetization in different phases exhibits distinct susceptibility to suppression by $\beta$.

Remarkably, DQPTs persist even in systems with strong initial spin polarization when the quench crosses a critical point (see Sec.~\ref{sec:DQPT}). The insensitivity of such DQPTs to the initial magnetization can be understood in terms of spin flips. Consider the XY model is initially in the $\mathrm{FM}_x$ phase with all spins polarized along the $x$-direction. If the system is quenched into the PM phase, the dominant term of the post-quench Hamiltonian becomes the transverse field along the $z$ direction. This $z$--oriented field flips spins from the initial $x$-polarized configuration, leading to the occurrence of a DQPT. Similarly, if the system is quenched into the $\mathrm{FM}_y$ phase, the post-quench Hamiltonian is dominated by spin-spin interactions along the $y$-direction. With a mean-field description, this is equivalent to each spin being subjected to an effective field in the $y$-direction. This $y$-oriented field also causes the $x$-polarized spins to flip, thereby triggering a DQPT.

\begin{figure}[tbhp] \centering%[tph]
\includegraphics[width=8.5cm]{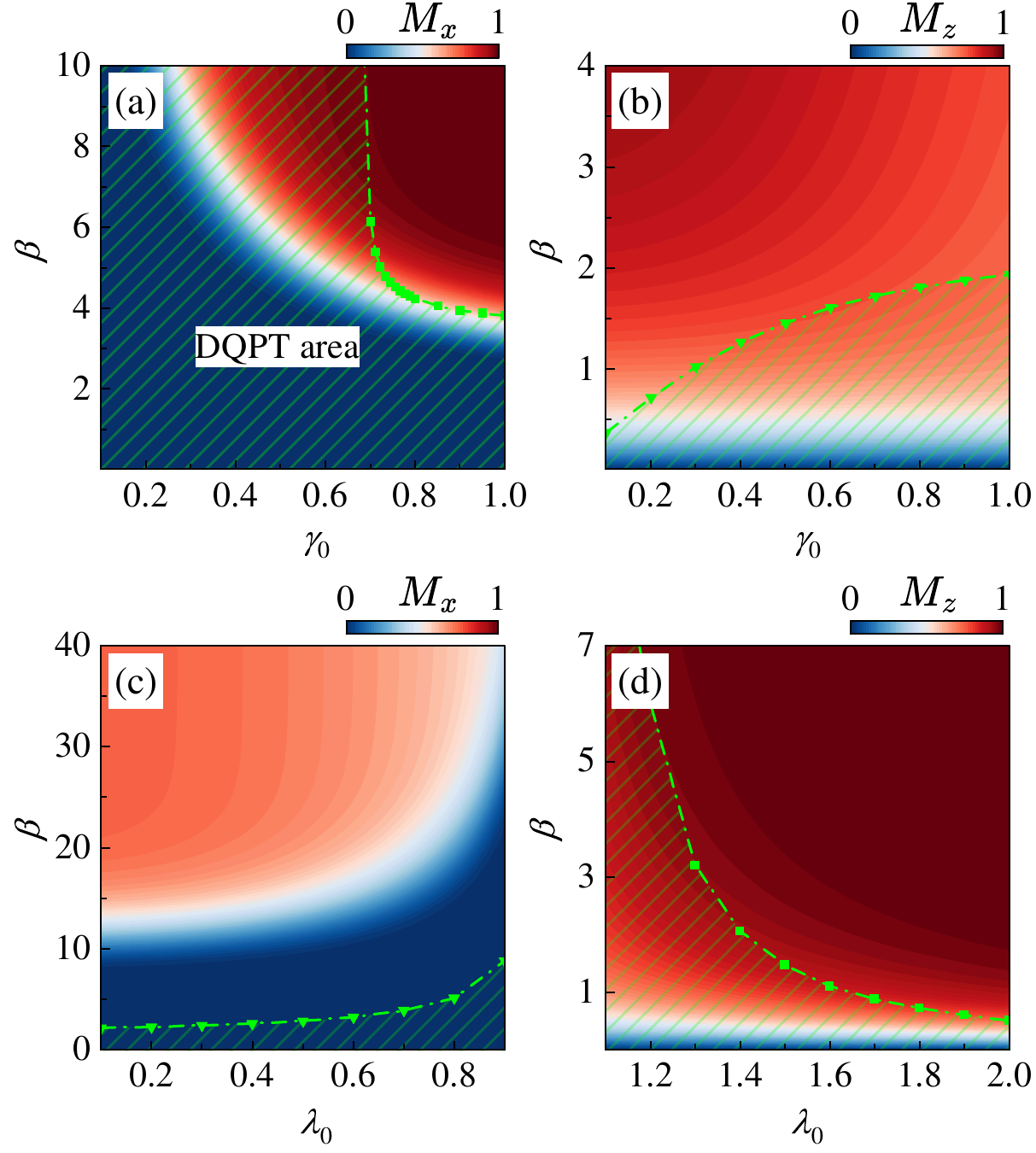}
\caption{Magnetization versus contour plots characterize the initial state $\psi(\lambda_0,\gamma_0,\beta)$, with the critical $\beta_c$ line (green dashed) marking the maximum intraphase quench amplitude. The hatched green region identifies the ``DQPT area'' where DQPT is accessible via the corresponding quench protocol.
(a) and (b) display the results for transverse field quenches, where $\lambda$ is quenched from 0 to 0.999 in the FM$_x$ phase and from 1.2 to 2 in the PM phase, respectively.
(c) and (d) show the corresponding behavior for anisotropy parameter quenches, with $\gamma$ quenched from 0.2 to 1 in the FM$_x$ phase and from 0.2 to $-1$ in the PM phase. Throughout, $\phi=-\pi/2$.}
\label{FIG5magnet}
\end{figure}

However, if the quench is performed within the same phase, the post-quench Hamiltonian alone cannot generally induce spin flips. In such a case, an additional mechanism should be introduced to drive a DQPT, such as the quantum coherence provided by the coherent Gibbs state in this paper. The ease of spin flipping thus determines the extent of additional resources needed for DQPT. We argue that this ease is controlled by the initial magnetization of the system: Larger magnetization impedes spin flips and raises the threshold for triggering a DQPT. In other words, there exists a profound and intimate connection between magnetization and DQPT. This relationship, detailed in Fig.~\ref{FIG5magnet}, is revealed through single-parameter intra-phase quenches, showing transverse field quenches within (a) the FM\(_x\) phase and (b) the PM phase, alongside anisotropy parameter quenches within (c) the FM\(_x\) phase and (d) the PM phase.

%%%%%%
%%%%%%
%%%%%%
%%%%%%

From the contour plot of magnetization in Fig.~\ref{FIG5magnet}, we can identify the magnetization corresponding to various initial states \((\lambda_0, \gamma_0, \beta)\). For each of these states, we perform an intra-phase quench on a single parameter, following either the protocol \((\lambda_0 \rightarrow \lambda_f, \gamma_0, \beta)\) or \((\lambda_0, \gamma_0 \rightarrow \gamma_f, \beta)\).
For a fixed quench protocol, a critical value $\beta_c$ can be found, below which a DQPT occurs. In general, the area beneath this $\beta_c$ curve defines the ``DQPT area'' for that quench, which tends to expand toward higher $\beta_c$ values as the quench amplitude increases (see Appendix~\ref{APDIX:quench amplitude} for details).
Since a larger \(\beta\) is associated with a higher magnetization, as shown in Fig.~\ref{FIG4temperatureM}, the expansion of the DQPT area implies that the upper magnetization limit for states capable of exhibiting a DQPT is raised within this area. Consequently, the limiting case of the maximum intra-phase quench amplitude yields the largest possible DQPT area. In Fig.~\ref{FIG5magnet}, we specifically mark this maximum $\beta_c$ curve (green dashed line) and its corresponding DQPT area.

We first analyze the intra-phase transverse field quench within the FM\(_x\) phase, with initial states parameterized by $(\lambda_0=0, \gamma_0, \beta)$, as shown in Fig.~\ref{FIG5magnet}(a).
For finite $\beta$, $M_x$ at fixed $\beta$ increases with $\gamma_0$ due to the enhanced spin-spin coupling along the $x$-direction, thereby forming a saturation region (dark red) in the upper-right corner of the contour plot. The ``DQPT area'' is defined by the maximum quench amplitude ($\lambda_f = 0.999$). A key finding is that initial states with zero magnetization (dark blue region) are universally susceptible to DQPTs. As the initial magnetization strengthens, a progressively larger quench amplitude is required to trigger a DQPT. The critical $\beta_c$ line (green dashed) for the maximum quench amplitude thus defines the boundary beyond which the initial magnetization is too strong to be overcome by any intra-phase transverse field quench. In essence, a weaker initial magnetization significantly facilitates the occurrence of a DQPT.

We now turn to the PM phase in Fig.~\ref{FIG5magnet}(b), where initial states are parameterized by $(\lambda_0=1.2, \gamma_0, \beta)$.
The response to a transverse field quench $(\lambda_0=1.2 \rightarrow \lambda_f)$ shows patterns consistent with the FM$_x$ case. The critical $\beta_c$ line for the maximum quench amplitude $\lambda_f = 2$ partitions the parameter space, with the DQPT area encompassing the zero magnetization region. This confirms that vanishing initial magnetization ensures susceptibility to DQPTs from transverse field quenches. Within the DQPT area, stronger initial magnetizations require larger quench amplitudes to trigger a DQPT, while states above the $\beta_c$ boundary possess magnetization too robust for any intra-phase transverse field quench to overcome.

We next investigate quenches of the anisotropy parameter. Fig.~\ref{FIG5magnet}(c) presents the magnetization contour plot for initial states within the FM$_x$ phase, parameterized as \((\lambda_0, \gamma_0=0.2, \beta)\).
However, the response of the system to an anisotropy parameter quench $(\gamma_0=0.2 \rightarrow \gamma_f=1)$ reveals a crucial distinction from previous cases. While the critical $\beta_c$ line again partitions the parameter space, the resulting DQPT area now fails to fully encompass the zero magnetization region. This indicates that certain initial states with zero magnetization cannot host a DQPT under this specific $\gamma$-quench protocol. However, systems prepared in these initial states can exhibit DQPTs under a transverse field quench (see Appendix~\ref{APDIX:FIG5c} for details), highlighting the protocol dependence of DQPTs. States outside the DQPT area and farther from the critical $\beta_c$ line are more strongly magnetized, making them less likely to undergo DQPTs under intra-phase $\gamma$-quenches.

Fig.~\ref{FIG5magnet}(d) completes our analysis by examining the PM phase under an anisotropy quench, with initial states $(\lambda_0, \gamma_0=0.2, \beta)$.
The behavior of the system under a maximum-amplitude anisotropy quench $(\gamma_0=0.2 \rightarrow \gamma_f=-1)$ reaffirms the established pattern. The resulting DQPT area fully contains the zero-magnetization region, confirming that sufficiently small initial magnetization guarantees susceptibility to DQPTs via $\gamma$-quenching. Within this area, triggering a DQPT requires progressively larger quench amplitudes as the initial magnetization increases. The critical $\beta_c$ line for the maximum quench amplitude again defines the fundamental limit beyond which no intra-phase $\gamma$-quench can induce a DQPT, regardless of amplitude.

Based on the preceding analysis, we establish that DQPTs within the same phase are closely related to the spin polarization of the system, with magnetization serving as a direct indicator. The essential mechanism of DQPTs fundamentally depends on whether the spins can be flipped under the post-quench dynamics. When the initial magnetization is zero, the system can readily undergo collective spin flips, leading to the emergence of a DQPT. In this specific case, there exists at least one type of quench under which a DQPT can occur. Within the ``DQPT area'' corresponding to the maximum quench amplitude, a larger initial magnetization requires a correspondingly larger quench amplitude to trigger a DQPT. This signifies that it becomes progressively more difficult to induce a DQPT via quenching as the initial magnetization increases. For initial states lying outside the maximum ``DQPT area'', their initial magnetization exceeds the maximum value observed within the DQPT area. Thus, for intra-phase quenches, the strong initial spin polarization of these states cannot be overcome even by the maximum quench amplitude, preventing the induction of a DQPT.

\section{Conclusion}
\label{sec:conclusion}
In summary, by taking the one-dimensional transverse-field XY model as an example, we discussed the quenching dynamics with coherent Gibbs states as starting points. We analytically obtain the profile of Fisher zeros, thereby determining the corresponding critical $\beta_c$. By comparing it with the initial magnetization, we obtained the relationship between the emergence of DQPT and the magnetization. The results reveal that under the condition of the same-phase quench, the occurrence of DQPT exhibits a significant dependence on the magnetization.
%%%%%%%%
That is, for a fixed maximum-amplitude quench protocol, a stronger initial magnetization within the corresponding DQPT area makes the emergence of DQPT more difficult. From the perspective of the physical mechanism, DQPTs originate from spin-flip dynamics. A weaker initial magnetization makes it easier for the system to overcome the initial spin polarization and undergo a collective spin flip during the quench, thereby facilitating a DQPT. Initial states that lie beyond this maximum DQPT area cannot undergo a DQPT, as their spin polarization resists the quench, even at the maximum amplitude.
%%%%%%%%
Experimentally, the DQPT has been realized in artificial systems such as cold atomic gases, superconducting circuits, etc. Meanwhile, detection techniques for observables like Lee-Yang zeros and magnetization have also been developed~\cite{PRL.2015.Peng,zhang2017observation}. Therefore, we hope that the theoretical predictions in this manuscript can bring benefits to related experiments and help uncover the physical mechanism of DQPT.
\acknowledgments
We sincerely thank X.-Y. Zhang and L.-F. Yu for helpful discussions. L.-Y. Luo, W.-L. Li and ZL are supported by the National Key Research and Development Program of China (Grant No.2022YFA1405300), Innovation Program for Quantum Science and Technology (Grant No. 2021ZD0301700), the Guangdong Basic and Applied Basic Research Foundation (Grant No.2021A1515012350), Guangdong Provincial Quantum Science Strategic Initiative(Grants No.GDZX2304002 and GDZX2404001), and the Open Fund of Key Laboratory of Atomic and Subatomic Structure and Quantum Control (Ministry of Education).

 L.-Y. Luo and W.-L. Li contribute equally to this work.

\appendix
\section{Magnetization}
\label{APDIX:Magnetization}
The magnetization along the $z$-direction is directly accessible as a local expectation value \cite{PRB.2024.miao}:
\begin{equation}
    M_z=\langle \sigma^z_j\rangle.
    \label{eq:Mz_def}
\end{equation}
Using the Jordan-Wigner transformation followed by a Fourier transformation, $M_z$ can be expressed as:
%%%
%%%
\begin{equation}\label{jw}
M_z=\frac{2}{N}\sum_{k}\langle c_{k}^{\dagger}c_{k}\rangle-1.
\end{equation}

In contrast, the magnetic order along the $\alpha$-direction ($\alpha = x, y$) is defined through the long-range limit of the corresponding spin-spin correlation function:
\begin{equation}M_{\alpha}=\sqrt{\lim_{r\to\infty}\langle \sigma^{\alpha}_i\sigma^{\alpha}_j \rangle},\end{equation}
where $r = |j - i|$. This definition is necessary because the expectation values $\langle \sigma^x \rangle$ and $\langle \sigma^y \rangle$ vanish directly due to the $Z_2$ symmetry of the system.

For notational clarity, we define the spin-spin correlation function $\langle \sigma^{\alpha}_i\sigma^{\alpha}_j \rangle$ as:
\begin{equation}
    C_{\alpha}(r)=\langle \sigma^{\alpha}_i\sigma^{\alpha}_j \rangle.
    \label{eq:corr_def}
\end{equation}
Through the Jordan-Wigner transformation, we can also express it by the product of $A_j=c^{\dagger}_j+c_j$ and $B_j=c^{\dagger}_j-c_j$ as
\begin{equation}
    \begin{aligned}
C_{x}(r) &= \left\langle(c_i - c_i^\dagger) \prod_{i < m < j} (1 - 2c_m^\dagger c_m) (c_j^\dagger + c_j) \right\rangle \\
&= \langle B_i A_{i+1} B_{i+1} \ldots A_{j-1} B_{j-1} A_j \rangle,
\label{EqCx}
\end{aligned}
\end{equation}
%%%
\begin{equation}
    C_{y}(r)=(-1)^{r}\langle A_iB_{i+1}A_{i+1}\ldots B_{j-1}A_{j-1}B_j\rangle.
\label{EqCy}
\end{equation}
The expectation values in Eqs.~(\ref{EqCx}) and (\ref{EqCy}) involve long strings of operators.
According to Wick's theorem, these can be written in terms of the Pfaffian of a skew-symmetric matrix, i.e.,
\begin{equation}C_x(r)=\mathrm{Pf}
\begin{vmatrix}
0 & G_1 & S_1 & G_2 & S_2 & \cdots & G_r \\
 & 0 & D_0 & Q_1 & D_1 & \cdots & Q_{r-1} \\
 & & 0 & G_1 & S_1 & \cdots & G_{r-1} \\
 & & & 0 & D_0 & \cdots & Q_{r-2} \\
 & & & & \ddots & \ddots & \vdots \\
 & & & & & 0 & G_1 \\
 & & & & & & 0
 \label{EQcorrelationx}
\end{vmatrix},\end{equation}
%%%
\begin{equation}\begin{aligned}
C_y(r)=(-1)^{r}\mathrm{Pf}
\begin{vmatrix}
0 & D_1 & Q_1 & D_2 & Q_2 & \cdots & D_r \\
 & 0 & G_0 & S_1 & G_1 & \cdots & S_{r-1} \\
 & & 0 & D_1 & Q_1 & \cdots & D_{r-1} \\
 & & & 0 & G_0 & \cdots & S_{r-2} \\
 & & & & \ddots & \ddots & \vdots \\
 & & & & & 0 & D_1 \\
 & & & & & & 0
\end{vmatrix}, \\
\label{EQcorrelationy}
\end{aligned}\end{equation}
%%%
where the matrix elements are the contractions of pairs defined as:
\begin{equation}
 \begin{aligned}
Q_r&=\langle A_iA_j\rangle\\
&=\sum_{k}\frac{-i\sin(kr)\langle c^{\dagger}_kc_k+c_{-k}c_k+c_k^{\dagger}c_{-k}^{\dagger}+c_{-k}c_{-k}^{\dagger}\rangle}{N},
\end{aligned}
\end{equation}
\begin{equation}
\begin{aligned}
    S_r&=\langle B_iB_j\rangle\\
    &=\sum_{k}\frac{i\sin{(kr)}\langle c_k^{\dagger}c_k-c_{-k}c_k-c_k^{\dagger}c_{-k}^{\dagger}+c_{-k}c_{-k}^{\dagger}\rangle}{N},
\end{aligned}
\end{equation}
\begin{equation}
  \begin{aligned}
G_r &= -D_{-r} \\
&= \langle B_iA_j \rangle \\
&= \sum_{k}\frac{\cos(kr) \langle c_k^{\dagger}c_k - c_{-k}c_{-k}^{\dagger} \rangle +i \sin(kr) \langle c_{-k}c_k - c_k^{\dagger}c_{-k}^{\dagger} \rangle}{N}.
\end{aligned}
\end{equation}

Finally, the fermionic two-operator expectation values for the coherent Gibbs initial state are given by:
\begin{equation}
\begin{aligned}
    &\langle c_k^{\dagger}c_{-k}^{\dagger}\rangle \\&=\frac{i(e^{-\beta\varepsilon}-e^{\beta\varepsilon})\sin{\theta_k}\cos{\theta_k}+e^{i\phi}\sin^2\theta_k+e^{-i\phi}\cos^2{\theta_k}}{Z_k},\\
&\langle c_k^{\dagger}c_{k}\rangle\\ & =\frac{i(e^{i\phi}-e^{-i\phi})\sin{\theta_k}\cos{\theta_k}+e^{\beta\varepsilon}\cos^2\theta_k+e^{-\beta\varepsilon}\sin^2{\theta_k}}{Z_k},\\
&\langle c_{-k}c_{k}\rangle \\&  =\frac{i(e^{\beta\varepsilon}-e^{-\beta\varepsilon})\sin{\theta_k}\cos{\theta_k}+e^{i\phi}\cos^2{\theta_k}+e^{-i\phi}\sin^2{\theta_k}}{Z_k},\\
&\langle c_{-k}c^{\dagger}_{-k}\rangle\\&=\frac{i(e^{-i\phi}-e^{i\phi})\sin{\theta_k}\cos{\theta_k}+e^{-\beta\varepsilon}\cos^2{\theta_k}+e^{\beta\varepsilon}\sin^2{\theta_k}}{Z_k},
\label{cc}
\end{aligned}
\end{equation}
where $Z_k= 2\cosh[\beta \varepsilon_k]$ is the associated partition function for the $k$-th mode.
%%%%%%
\section{The effect of the quench amplitude on the ``DQPT area"}
\label{APDIX:quench amplitude}

\begin{figure}[tbhp] \centering%[tph]
\includegraphics[width=8.5cm]{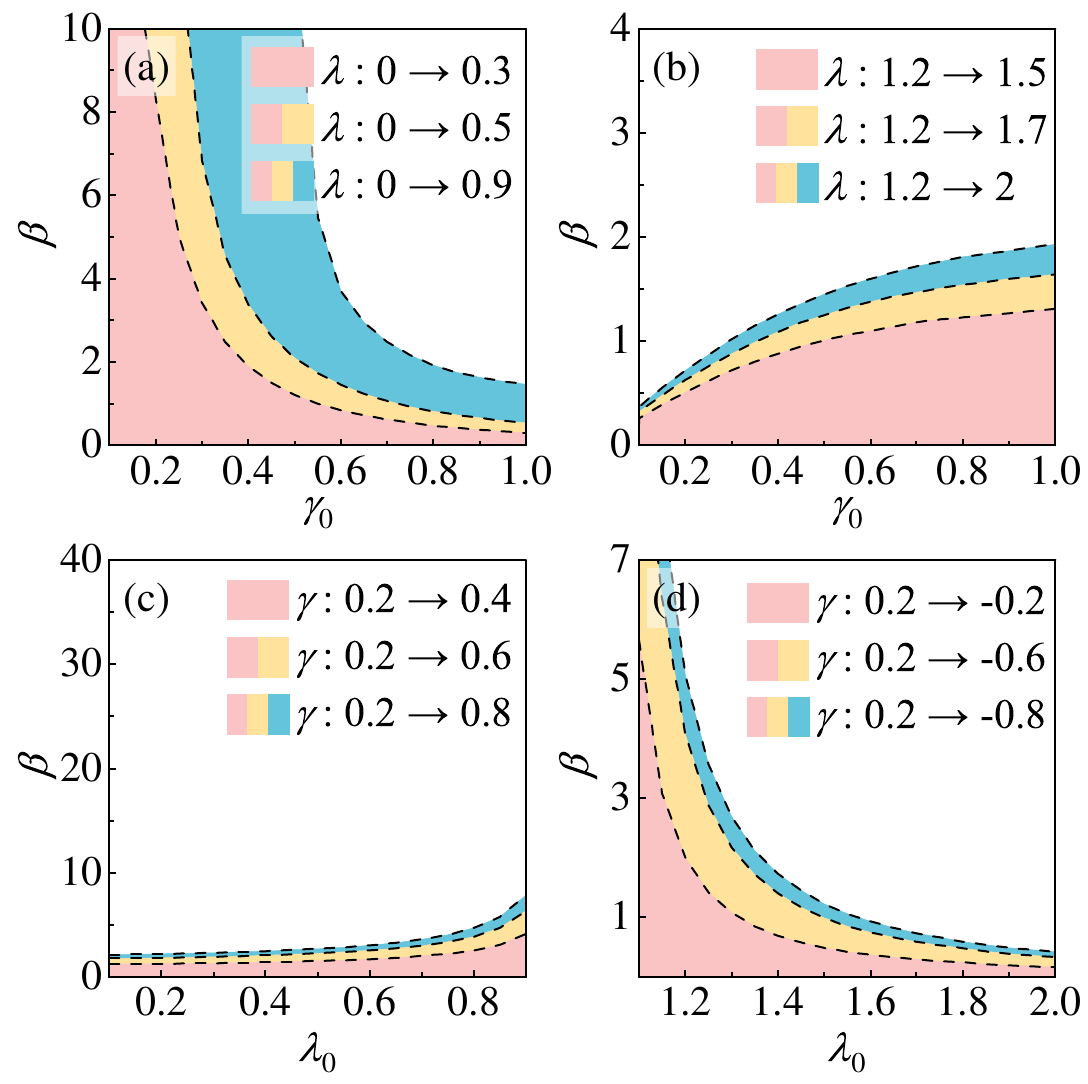}
\caption{Influence of quench amplitude on the critical $\beta_c$ for DQPTs. Dashed lines represent the critical $\beta_c$ at different quench amplitudes, with the corresponding ``DQPT area'' shaded in matching colors. Transverse-field quenches are performed in the region of FM$_x$ (a) and PM (b) phases. (c) and (d) represent quenches of the anisotropy parameter within the FM$_x$ and PM phases, respectively. Throughout, $\phi=-\pi/2$.}
\label{fig:quench_amplitude}
\end{figure}

We investigate the influence of quench amplitude on the critical $\beta_c$ for DQPTs across various intraphase quenches in Fig.~\ref{fig:quench_amplitude}. For each quench amplitude, we plot the corresponding critical $\beta_c$ line as a dashed curve. The region where a DQPT can occur under each specific quench protocol, termed the ``DQPT area'', is colored accordingly.

%%%%%%
%%%%%%
%%%%%%
%%%%%%

A universal trend emerges from these results, applicable to both transverse field and anisotropy parameter quenches within either the FM$_x$ or PM phases. When the pre-quench parameters are fixed, an increase in the quench amplitude systematically shifts the critical $\beta_c$ line to higher values. This shift, in turn, causes a progressive expansion of the ``DQPT area''.
Taking the transverse field quench ($\lambda_0=0 \rightarrow \lambda_f$) in the FM$_x$ phase as a concrete example [see Fig.~\ref{fig:quench_amplitude}(a)], for a fixed $\gamma_0$, a larger quench amplitude results in a higher critical $\beta_c$. Given the established positive correlation between magnetization and $\beta$ from Fig.~\ref{FIG4temperatureM}, a higher critical $\beta_c$ directly implies that initial states with stronger magnetization can now host a DQPT under the larger quench.

\section{More discussion about the DQPTs in Fig.~\ref{FIG5magnet}(c)}
\label{APDIX:FIG5c}
To further characterize the non-equilibrium dynamics, we calculate the so-called rate function, which plays a role analogous to the free energy density in equilibrium statistical mechanics:
%%%
%%%
\begin{equation}
r\left(t\right)=-\lim_{N\to\infty}\frac1N\ln\mathcal{L}\left(t\right),
\end{equation}
%%%
%%%
where $\mathcal{L}\left(t\right)=\left|\mathcal{G}\left(t\right)\right|^2$ is the Loschmidt echo, quantifying the probability of the system returning to its initial state after time evolution. Notably, cusp-like singularities at critical times in the rate function $r\left(t\right)$, reflecting nonanalytic behavior, are regarded as hallmarks of DQPTs. In the thermodynamic limit, the rate function \( r(t) \) admits the following analytical expression:
%%%
%%%
\begin{widetext}
\begin{equation}
r(t) = -\frac{1}{\pi} \mathrm{Re} \left[ \int_0^\pi dk \,
\ln \left\{
\cos[\varepsilon_k(\gamma',\lambda') t] +
i \sin[\varepsilon_k(\gamma',\lambda') t] \left[
\cos(2\Delta_{\theta_k}) \tanh(\beta \varepsilon_k(\gamma,\lambda))
- \frac{2}{Z_k(\gamma,\lambda)} \sin\phi \sin(2\Delta_{\theta_k})
\right]
\right\}
\right].
\label{r_t}
\end{equation}
\end{widetext}
%%%
%%%

%%%%%%
%%%%%%
\begin{figure}[tbhp] \centering%[tph]
\includegraphics[width=5.5cm]{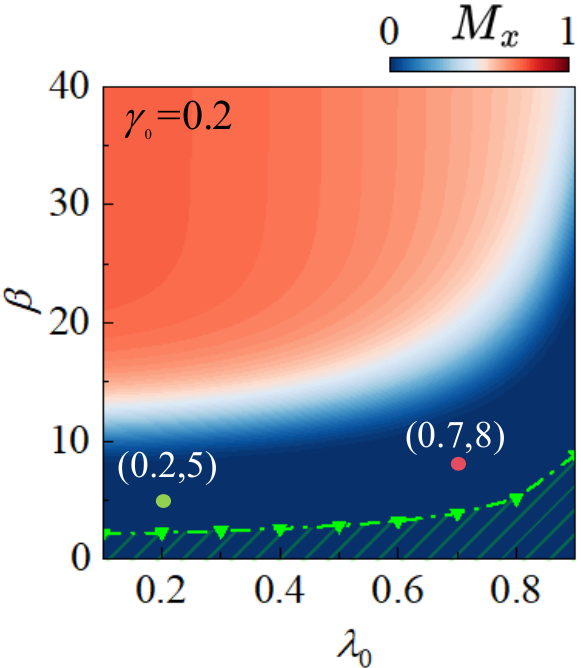}
\caption{Magnetization contour plot in the $\mathrm{FM}_{x}$ phase for initial conditions $(\lambda_0,\gamma_0=0.2,\beta)$.
The green curve marks the critical $\beta_c$ line at the maximum anisotropy parameter quench amplitude ($\gamma_0=0.2 \rightarrow \gamma_f=1$), with the corresponding ``DQPT area'' indicated by green hatching.
Two specific initial states with zero magnetization that lie outside the DQPT area are highlighted: $(\lambda_0=0.2,\gamma_0=0.2,\beta=5)$ (green dot) and $(\lambda_0=0.7,\gamma_0=0.2,\beta=8)$ (magenta dot). Throughout, $\phi=-\pi/2$.}
\label{Fig5(c)}
\end{figure}
%%%%%%
%%%%%%

%%%%%%
%%%%%%
\begin{figure}[tbhp] \centering%[tph]
\includegraphics[width=8.5cm]{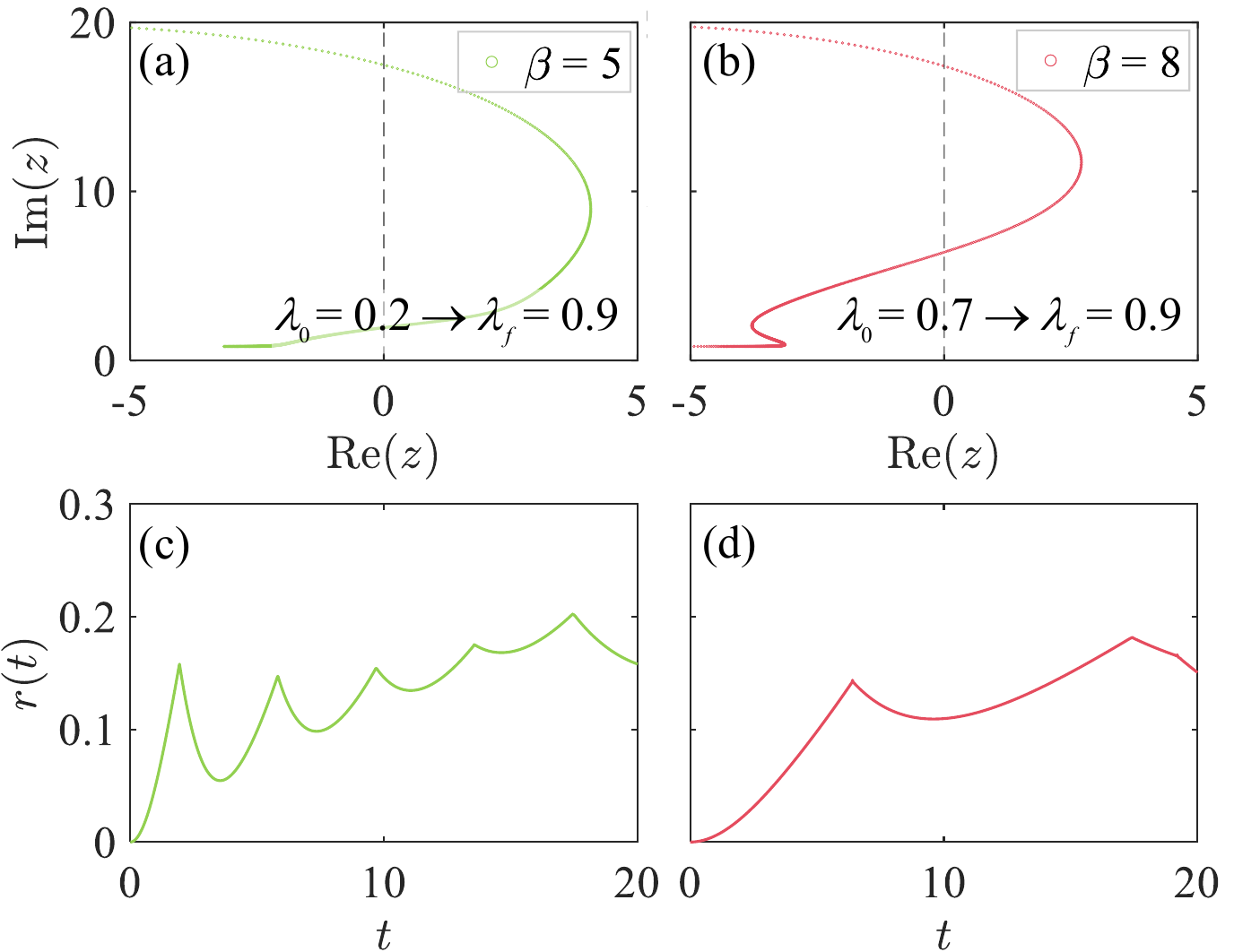}
\caption{DQPT signatures for the states marked in Fig.~\ref{Fig5(c)} after a transverse field quench to $\lambda_f=0.9$. (a) and (b) show the Fisher zeros for the green and magenta dots, respectively. (c) and (d) display the corresponding rate functions. Throughout, $\phi=-\pi/2$.}
\label{FigDQPTin5c}
\end{figure}
%%%%%%
%%%%%%

To elucidate the protocol-dependent nature of DQPTs, we examine two specially marked initial states in Fig.~\ref{Fig5(c)}. These states possess zero magnetization but lie outside the ``DQPT area'' for the anisotropy parameter quench. We now investigate these particular initial conditions by performing a transverse field quench and monitoring the subsequent dynamics through Fisher zeros and the rate function.
When the transverse field is quenched to $0.9$, the lines of Fisher zeros cross the imaginary axis in both cases [Fig.~\ref{FigDQPTin5c}(a) and (b)]. Concurrently, the rate functions develop non-analytic cusps [Fig.~\ref{FigDQPTin5c}(c) and (d)]. This phenomenon demonstrates that for a system with zero initial magnetization, a DQPT can be induced by at least one of the two quench protocols: Either a transverse-field quench or a $\gamma$-quench. This underscores that dynamical criticality is not solely an intrinsic property of the initial state but is fundamentally protocol-dependent.

\newpage
\bibliography{main}

@article{eisert2015quantum,
  title = {Quantum Many-Body Systems out of Equilibrium},
  author = {Eisert, J. and Friesdorf, M. and Gogolin, C.},
  year = {2015},
  month = feb,
  journal = {Nature Physics},
  volume = {11},
  number = {2},
  pages = {124--130},
  issn = {1745-2481},
  doi = {10.1038/nphys3215},
}

@article{RevModPhys.2011.Polkovnikov,
  title = {Colloquium: Nonequilibrium dynamics of closed interacting quantum systems},
  author = {Polkovnikov, Anatoli and Sengupta, Krishnendu and Silva, Alessandro and Vengalattore, Mukund},
  journal = {Rev. Mod. Phys.},
  volume = {83},
  issue = {3},
  pages = {863--883},
  numpages = {0},
  year = {2011},
  month = {Aug},
  publisher = {American Physical Society},
  doi = {10.1103/RevModPhys.83.863},
  url = {https://link.aps.org/doi/10.1103/RevModPhys.83.863}
}

@article{PRL.2005.Zurek,
  title = {Dynamics of a Quantum Phase Transition},
  author = {Zurek, Wojciech H. and Dorner, Uwe and Zoller, Peter},
  journal = {Phys. Rev. Lett.},
  volume = {95},
  issue = {10},
  pages = {105701},
  numpages = {4},
  year = {2005},
  month = {Sep},
  publisher = {American Physical Society},
  doi = {10.1103/PhysRevLett.95.105701},
  url = {https://link.aps.org/doi/10.1103/PhysRevLett.95.105701}
}

@article{marino2022dynamical,
  title = {Dynamical Phase Transitions in the Collisionless Pre-Thermal States of Isolated Quantum Systems: Theory and Experiments},
  author = {Marino, Jamir and Eckstein, Martin and Foster, Matthew S and Rey, Ana Maria},
  year = {2022},
  month = oct,
  journal = {Reports on Progress in Physics},
  volume = {85},
  number = {11},
  pages = {116001},
  publisher = {IOP Publishing},
  doi = {10.1088/1361-6633/ac906c},
}

@article{heyl2019dynamical,
  title = {Dynamical Quantum Phase Transitions: {{A}} Brief Survey},
  author = {Heyl, Markus},
  year = {2019},
  month = feb,
  journal = {Europhysics Letters},
  volume = {125},
  number = {2},
  pages = {26001},
  publisher = {{EDP Sciences, IOP Publishing and Societ{\`a} Italiana di Fisica}},
  doi = {10.1209/0295-5075/125/26001},
}

@article{heyl2018dynamical,
  title = {Dynamical Quantum Phase Transitions: A Review},
  shorttitle = {Dynamical Quantum Phase Transitions},
  author = {Heyl, Markus},
  year = {2018},
  month = may,
  journal = {Reports on Progress in Physics},
  volume = {81},
  number = {5},
  pages = {054001},
  issn = {0034-4885, 1361-6633},
  doi = {10.1088/1361-6633/aaaf9a},
  urldate = {2025-01-11},
  langid = {english}
}

@article{zvyagin2016dynamical,
  title = {Dynamical Quantum Phase Transitions ({{Review Article}})},
  author = {Zvyagin, A. A.},
  year = {2016},
  month = nov,
  journal = {Low Temperature Physics},
  volume = {42},
  number = {11},
  pages = {971--994},
  issn = {1063-777X, 1090-6517},
  doi = {10.1063/1.4969869},
  urldate = {2025-04-22},
  langid = {english}
}

@article{eckstein2008nonthermal,
  title = {Nonthermal Steady States after an Interaction Quench in the Falicov-Kimball Model},
  author = {Eckstein, Martin and Kollar, Marcus},
  journal = {Phys. Rev. Lett.},
  volume = {100},
  issue = {12},
  pages = {120404},
  numpages = {4},
  year = {2008},
  month = {Mar},
  publisher = {American Physical Society},
  doi = {10.1103/PhysRevLett.100.120404},
  url = {https://link.aps.org/doi/10.1103/PhysRevLett.100.120404}
}

@article{eckstein2009thermalization,
  title = {Thermalization after an Interaction Quench in the Hubbard Model},
  author = {Eckstein, Martin and Kollar, Marcus and Werner, Philipp},
  journal = {Phys. Rev. Lett.},
  volume = {103},
  issue = {5},
  pages = {056403},
  numpages = {4},
  year = {2009},
  month = {Jul},
  publisher = {American Physical Society},
  doi = {10.1103/PhysRevLett.103.056403},
  url = {https://link.aps.org/doi/10.1103/PhysRevLett.103.056403}
}

@article{moeckel2008interaction,
  title = {Interaction Quench in the Hubbard Model},
  author = {Moeckel, Michael and Kehrein, Stefan},
  journal = {Phys. Rev. Lett.},
  volume = {100},
  issue = {17},
  pages = {175702},
  numpages = {4},
  year = {2008},
  month = {May},
  publisher = {American Physical Society},
  doi = {10.1103/PhysRevLett.100.175702},
  url = {https://link.aps.org/doi/10.1103/PhysRevLett.100.175702}
}

@article{moeckel2010crossover,
  title = {Crossover from Adiabatic to Sudden Interaction Quenches in the {{Hubbard}} Model: Prethermalization and Non-Equilibrium Dynamics},
  author = {Moeckel, Michael and Kehrein, Stefan},
  year = {2010},
  month = may,
  journal = {New Journal of Physics},
  volume = {12},
  number = {5},
  pages = {055016},
  doi = {10.1088/1367-2630/12/5/055016},
}

@article{Sciolla2010Infinite-Dimensional,
  title = {Quantum Quenches and Off-Equilibrium Dynamical Transition in the Infinite-Dimensional Bose-Hubbard Model},
  author = {Sciolla, Bruno and Biroli, Giulio},
  journal = {Phys. Rev. Lett.},
  volume = {105},
  issue = {22},
  pages = {220401},
  numpages = {4},
  year = {2010},
  month = {Nov},
  publisher = {American Physical Society},
  doi = {10.1103/PhysRevLett.105.220401},
  url = {https://link.aps.org/doi/10.1103/PhysRevLett.105.220401}
}

@article{Sciolla_2013,
  title = {Quantum quenches, dynamical transitions, and off-equilibrium quantum criticality},
  author = {Sciolla, Bruno and Biroli, Giulio},
  journal = {Phys. Rev. B},
  volume = {88},
  issue = {20},
  pages = {201110},
  numpages = {5},
  year = {2013},
  month = {Nov},
  publisher = {American Physical Society},
  doi = {10.1103/PhysRevB.88.201110},
  url = {https://link.aps.org/doi/10.1103/PhysRevB.88.201110}
}

@article{gambassi2011quantum,
  title = {Quantum Quenches as Classical Critical Films},
  author = {Gambassi, A. and Calabrese, P.},
  year = {2011},
  month = sep,
  journal = {Europhysics Letters},
  volume = {95},
  number = {6},
  pages = {66007},
  doi = {10.1209/0295-5075/95/66007},
}

@article{Maraga2015aging,
  title = {Aging and coarsening in isolated quantum systems after a quench: Exact results for the quantum $\text{O}(N)$ model with $N$ $\ensuremath{\rightarrow}$ $\ensuremath{\infty}$},
  author = {Maraga, Anna and Chiocchetta, Alessio and Mitra, Aditi and Gambassi, Andrea},
  journal = {Phys. Rev. E},
  volume = {92},
  issue = {4},
  pages = {042151},
  numpages = {14},
  year = {2015},
  month = {Oct},
  publisher = {American Physical Society},
  doi = {10.1103/PhysRevE.92.042151},
  url = {https://link.aps.org/doi/10.1103/PhysRevE.92.042151}
}

@article{chandran2013equilibration,
  title = {Equilibration and coarsening in the quantum $O(N)$ model at infinite $N$},
  author = {Chandran, Anushya and Nanduri, Arun and Gubser, S. S. and Sondhi, S. L.},
  journal = {Phys. Rev. B},
  volume = {88},
  issue = {2},
  pages = {024306},
  numpages = {10},
  year = {2013},
  month = {Jul},
  publisher = {American Physical Society},
  doi = {10.1103/PhysRevB.88.024306},
  url = {https://link.aps.org/doi/10.1103/PhysRevB.88.024306}
}

@article{smacchia2015exploring,
  title = {Exploring dynamical phase transitions and prethermalization with quantum noise of excitations},
  author = {Smacchia, Pietro and Knap, Michael and Demler, Eugene and Silva, Alessandro},
  journal = {Phys. Rev. B},
  volume = {91},
  issue = {20},
  pages = {205136},
  numpages = {9},
  year = {2015},
  month = {May},
  publisher = {American Physical Society},
  doi = {10.1103/PhysRevB.91.205136},
  url = {https://link.aps.org/doi/10.1103/PhysRevB.91.205136}
}

@article{halimeh2017prethermalization,
  title = {Prethermalization and persistent order in the absence of a thermal phase transition},
  author = {Halimeh, Jad C. and Zauner-Stauber, Valentin and McCulloch, Ian P. and de Vega, In\'es and Schollw\"ock, Ulrich and Kastner, Michael},
  journal = {Phys. Rev. B},
  volume = {95},
  issue = {2},
  pages = {024302},
  numpages = {7},
  year = {2017},
  month = {Jan},
  publisher = {American Physical Society},
  doi = {10.1103/PhysRevB.95.024302},
  url = {https://link.aps.org/doi/10.1103/PhysRevB.95.024302}
}

@article{mori2018thermalization,
  title = {Thermalization and Prethermalization in Isolated Quantum Systems: A Theoretical Overview},
  author = {Mori, Takashi and Ikeda, Tatsuhiko N and Kaminishi, Eriko and Ueda, Masahito},
  year = {2018},
  month = may,
  journal = {Journal of Physics B: Atomic, Molecular and Optical Physics},
  volume = {51},
  number = {11},
  pages = {112001},
  publisher = {IOP Publishing},
  doi = {10.1088/1361-6455/aabcdf},
}

@article{zhang2017observation,
  title = {Observation of a Many-Body Dynamical Phase Transition with a 53-Qubit Quantum Simulator},
  author = {Zhang, J. and Pagano, G. and Hess, P. W. and Kyprianidis, A. and Becker, P. and Kaplan, H. and Gorshkov, A. V. and Gong, Z.-X. and Monroe, C.},
  year = {2017},
  month = nov,
  journal = {Nature},
  volume = {551},
  number = {7682},
  pages = {601--604},
  publisher = {{Springer Science and Business Media LLC}},
  issn = {0028-0836, 1476-4687},
  doi = {10.1038/nature24654},
  urldate = {2025-01-14},
  copyright = {http://www.springer.com/tdm},
  langid = {english},
}

@article{muniz2020exploring,
  title = {Exploring Dynamical Phase Transitions with Cold Atoms in an Optical Cavity},
  author = {Muniz, Juan A. and Barberena, Diego and {Lewis-Swan}, Robert J. and Young, Dylan J. and Cline, Julia R. K. and Rey, Ana Maria and Thompson, James K.},
  year = {2020},
  month = apr,
  journal = {Nature},
  volume = {580},
  number = {7805},
  pages = {602--607},
  issn = {0028-0836, 1476-4687},
  doi = {10.1038/s41586-020-2224-x},
  urldate = {2025-09-01},
  langid = {english},
}

@article{
smale2019observation,
author = {Scott Smale  and Peiru He  and Ben A. Olsen  and Kenneth G. Jackson  and Haille Sharum  and Stefan Trotzky  and Jamir Marino  and Ana Maria Rey  and Joseph H. Thywissen },
title = {Observation of a transition between dynamical phases in a quantum degenerate Fermi gas},
journal = {Science Advances},
volume = {5},
number = {8},
pages = {eaax1568},
year = {2019},
doi = {10.1126/sciadv.aax1568},
URL = {https://www.science.org/doi/abs/10.1126/sciadv.aax1568},
}

@article{tian2020observation,
  title = {Observation of Dynamical Quantum Phase Transitions with Correspondence in an Excited State Phase Diagram},
  author = {Tian, T. and Yang, H.-X. and Qiu, L.-Y. and Liang, H.-Y. and Yang, Y.-B. and Xu, Y. and Duan, L.-M.},
  journal = {Phys. Rev. Lett.},
  volume = {124},
  issue = {4},
  pages = {043001},
  numpages = {6},
  year = {2020},
  month = {Jan},
  publisher = {American Physical Society},
  doi = {10.1103/PhysRevLett.124.043001},
  url = {https://link.aps.org/doi/10.1103/PhysRevLett.124.043001}
}

@article{PRL.2013.heyl,
  title = {Dynamical Quantum Phase Transitions in the Transverse-Field Ising Model},
  author = {Heyl, M. and Polkovnikov, A. and Kehrein, S.},
  journal = {Phys. Rev. Lett.},
  volume = {110},
  issue = {13},
  pages = {135704},
  numpages = {5},
  year = {2013},
  month = {Mar},
  publisher = {American Physical Society},
  doi = {10.1103/PhysRevLett.110.135704},
  url = {https://link.aps.org/doi/10.1103/PhysRevLett.110.135704}
}

@article{PRL.2014.heyl,
  title = {Dynamical Quantum Phase Transitions in Systems with Broken-Symmetry Phases},
  author = {Heyl, M.},
  journal = {Phys. Rev. Lett.},
  volume = {113},
  issue = {20},
  pages = {205701},
  numpages = {5},
  year = {2014},
  month = {Nov},
  publisher = {American Physical Society},
  doi = {10.1103/PhysRevLett.113.205701},
  url = {https://link.aps.org/doi/10.1103/PhysRevLett.113.205701}
}

@article{PRB.2014.Vajna,
  title = {Disentangling dynamical phase transitions from equilibrium phase transitions},
  author = {Vajna, Szabolcs and D\'ora, Bal\'azs},
  journal = {Phys. Rev. B},
  volume = {89},
  issue = {16},
  pages = {161105},
  numpages = {5},
  year = {2014},
  month = {Apr},
  publisher = {American Physical Society},
  doi = {10.1103/PhysRevB.89.161105},
  url = {https://link.aps.org/doi/10.1103/PhysRevB.89.161105}
}

@article{PRL.2015.hely,
  title = {Scaling and Universality at Dynamical Quantum Phase Transitions},
  author = {Heyl, Markus},
  journal = {Phys. Rev. Lett.},
  volume = {115},
  issue = {14},
  pages = {140602},
  numpages = {5},
  year = {2015},
  month = {Oct},
  publisher = {American Physical Society},
  doi = {10.1103/PhysRevLett.115.140602},
  url = {https://link.aps.org/doi/10.1103/PhysRevLett.115.140602}
}

@article{PRB.2016.Budich,
  title = {Dynamical topological order parameters far from equilibrium},
  author = {Budich, Jan Carl and Heyl, Markus},
  journal = {Phys. Rev. B},
  volume = {93},
  issue = {8},
  pages = {085416},
  numpages = {7},
  year = {2016},
  month = {Feb},
  publisher = {American Physical Society},
  doi = {10.1103/PhysRevB.93.085416},
  url = {https://link.aps.org/doi/10.1103/PhysRevB.93.085416}
}

@article{PRB.2017.Bhattacharya,
  title = {Mixed state dynamical quantum phase transitions},
  author = {Bhattacharya, Utso and Bandyopadhyay, Souvik and Dutta, Amit},
  journal = {Phys. Rev. B},
  volume = {96},
  issue = {18},
  pages = {180303},
  numpages = {5},
  year = {2017},
  month = {Nov},
  publisher = {American Physical Society},
  doi = {10.1103/PhysRevB.96.180303},
  url = {https://link.aps.org/doi/10.1103/PhysRevB.96.180303}
}

@article{PRB.2017.heyl,
  title = {Dynamical topological quantum phase transitions for mixed states},
  author = {Heyl, M. and Budich, J. C.},
  journal = {Phys. Rev. B},
  volume = {96},
  issue = {18},
  pages = {180304},
  numpages = {5},
  year = {2017},
  month = {Nov},
  publisher = {American Physical Society},
  doi = {10.1103/PhysRevB.96.180304},
  url = {https://link.aps.org/doi/10.1103/PhysRevB.96.180304}
}

@article{PRB.2020.cao,
  title = {Influence of weak disorder on the dynamical quantum phase transitions in the anisotropic XY chain},
  author = {Cao, Kaiyuan and Li, Wenwen and Zhong, Ming and Tong, Peiqing},
  journal = {Phys. Rev. B},
  volume = {102},
  issue = {1},
  pages = {014207},
  numpages = {10},
  year = {2020},
  month = {Jul},
  publisher = {American Physical Society},
  doi = {10.1103/PhysRevB.102.014207},
  url = {https://link.aps.org/doi/10.1103/PhysRevB.102.014207}
}

@article{arxiv.2024.cao,
   title={Aperiodic dynamical quantum phase transition in multi-band Bloch Hamiltonian and its origin},
   volume={36},
   ISSN={1361-648X},
   url={http://dx.doi.org/10.1088/1361-648X/ad1a5a},
   DOI={10.1088/1361-648x/ad1a5a},
   number={15},
   journal={Journal of Physics: Condensed Matter},
   publisher={IOP Publishing},
   author={Cao, Kaiyuan and Guo, Hao and Yang, Guangwen},
   year={2024},
   month=jan, pages={155401} }

@article{NJP.2023.Cheraghi,
   title={Dynamical quantum phase transitions following double quenches: persistence of the initial state vs dynamical phases},
   volume={25},
   ISSN={1367-2630},
   url={http://dx.doi.org/10.1088/1367-2630/ad016e},
   DOI={10.1088/1367-2630/ad016e},
   number={10},
   journal={New Journal of Physics},
   publisher={IOP Publishing},
   author={Cheraghi, Hadi and Sedlmayr, Nicholas},
   year={2023},
   month=oct, pages={103035} }

@article{ScientificReports.2019.Jafari,
  title={Dynamical Quantum Phase Transition and Quasi Particle Excitation},
  author={R. Jafari},
  journal={Scientific Reports},
  year={2019},
  volume={9},
  url={https://api.semanticscholar.org/CorpusID:67861874}
}

@article{fisher1965boulder,
  title={Boulder Lectures in Theoretical Physics vol. 7},
  author={Fisher, ME},
  journal={Colarado: University of Colorado},
  year={1965}
}

@article{PRL.2018.LANG,
  title = {Dynamical Quantum Phase Transitions: A Geometric Picture},
  author = {Lang, Johannes and Frank, Bernhard and Halimeh, Jad C.},
  journal = {Phys. Rev. Lett.},
  volume = {121},
  issue = {13},
  pages = {130603},
  numpages = {6},
  year = {2018},
  month = {Sep},
  publisher = {American Physical Society},
  doi = {10.1103/PhysRevLett.121.130603},
  url = {https://link.aps.org/doi/10.1103/PhysRevLett.121.130603}
}

@article{PRL.2018.Zunkovic,
  title = {Dynamical Quantum Phase Transitions in Spin Chains with Long-Range Interactions: Merging Different Concepts of Nonequilibrium Criticality},
  author = {\ifmmode \check{Z}\else \v{Z}\fi{}unkovi\ifmmode \check{c}\else \v{c}\fi{}, Bojan and Heyl, Markus and Knap, Michael and Silva, Alessandro},
  journal = {Phys. Rev. Lett.},
  volume = {120},
  issue = {13},
  pages = {130601},
  numpages = {6},
  year = {2018},
  month = {Mar},
  publisher = {American Physical Society},
  doi = {10.1103/PhysRevLett.120.130601},
  url = {https://link.aps.org/doi/10.1103/PhysRevLett.120.130601}
}

@article{PRB.2017.Halimeh,
  title = {Dynamical phase diagram of quantum spin chains with long-range interactions},
  author = {Halimeh, Jad C. and Zauner-Stauber, Valentin},
  journal = {Phys. Rev. B},
  volume = {96},
  issue = {13},
  pages = {134427},
  numpages = {5},
  year = {2017},
  month = {Oct},
  publisher = {American Physical Society},
  doi = {10.1103/PhysRevB.96.134427},
  url = {https://link.aps.org/doi/10.1103/PhysRevB.96.134427}
}

@article{PRE.2017.Zauner,
  title = {Probing the anomalous dynamical phase in long-range quantum spin chains through Fisher-zero lines},
  author = {Zauner-Stauber, Valentin and Halimeh, Jad C.},
  journal = {Phys. Rev. E},
  volume = {96},
  issue = {6},
  pages = {062118},
  numpages = {8},
  year = {2017},
  month = {Dec},
  publisher = {American Physical Society},
  doi = {10.1103/PhysRevE.96.062118},
  url = {https://link.aps.org/doi/10.1103/PhysRevE.96.062118}
}

@article{PRE.2016.Divakaran,
  title = {Tuning the presence of dynamical phase transitions in a generalized $XY$ spin chain},
  author = {Divakaran, Uma and Sharma, Shraddha and Dutta, Amit},
  journal = {Phys. Rev. E},
  volume = {93},
  issue = {5},
  pages = {052133},
  numpages = {7},
  year = {2016},
  month = {May},
  publisher = {American Physical Society},
  doi = {10.1103/PhysRevE.93.052133},
  url = {https://link.aps.org/doi/10.1103/PhysRevE.93.052133}
}

@article{CPB.2022.cao,
  title = {Dynamical Quantum Phase Transition in {{XY}} Chains with the {{Dzyaloshinskii-Moriya}} and {{XZY}}--{{YZX}} Three-Site Interactions},
  author = {Cao, Kaiyuan and Zhong, Ming and Tong, Peiqing},
  year = {2022},
  month = may,
  journal = {Chinese Physics B},
  volume = {31},
  number = {6},
  pages = {060505},
  issn = {1674-1056},
  doi = {10.1088/1674-1056/ac4a6e},
  urldate = {2025-09-01},
}

@article{ScientificReports.2020.Porta,
  title = {Topological Classification of Dynamical Quantum Phase Transitions in the Xy Chain},
  author = {Porta, Sergio and Cavaliere, Fabio and Sassetti, Maura and Traverso Ziani, Niccol{\`o}},
  year = {2020},
  month = jul,
  journal = {Scientific Reports},
  volume = {10},
  number = {1},
  pages = {12766},
  issn = {2045-2322},
  doi = {10.1038/s41598-020-69621-8},
}

@article{PRB.2025.ZENG,
  title = {Relation between equilibrium quantum phase transitions and dynamical quantum phase transitions in two-band systems},
  author = {Zeng, Yumeng and Chen, Shu},
  journal = {Phys. Rev. B},
  volume = {112},
  issue = {6},
  pages = {064307},
  numpages = {9},
  year = {2025},
  month = {Aug},
  publisher = {American Physical Society},
  doi = {10.1103/srx7-cpl4},
  url = {https://link.aps.org/doi/10.1103/srx7-cpl4}
}

@article{PRB.2015.Vajna,
  title = {Topological classification of dynamical phase transitions},
  author = {Vajna, Szabolcs and D\'ora, Bal\'azs},
  journal = {Phys. Rev. B},
  volume = {91},
  issue = {15},
  pages = {155127},
  numpages = {5},
  year = {2015},
  month = {Apr},
  publisher = {American Physical Society},
  doi = {10.1103/PhysRevB.91.155127},
  url = {https://link.aps.org/doi/10.1103/PhysRevB.91.155127}
}

@article{PRB.2015.Schmitt,
  title = {Dynamical quantum phase transitions in the Kitaev honeycomb model},
  author = {Schmitt, Markus and Kehrein, Stefan},
  journal = {Phys. Rev. B},
  volume = {92},
  issue = {7},
  pages = {075114},
  numpages = {13},
  year = {2015},
  month = {Aug},
  publisher = {American Physical Society},
  doi = {10.1103/PhysRevB.92.075114},
  url = {https://link.aps.org/doi/10.1103/PhysRevB.92.075114}
}

@article{JPA.2016.Jafari,
   title={Quench dynamics and ground state fidelity of the one-dimensional extended quantum compass model in a transverse field},
   volume={49},
   ISSN={1751-8121},
   url={http://dx.doi.org/10.1088/1751-8113/49/18/185004},
   DOI={10.1088/1751-8113/49/18/185004},
   number={18},
   journal={Journal of Physics A: Mathematical and Theoretical},
   publisher={IOP Publishing},
   author={Jafari, R},
   year={2016},
   month=apr, pages={185004} }

@article{PRL.2017.Jafari,
  title = {Loschmidt Echo Revivals: Critical and Noncritical},
  author = {Jafari, R. and Johannesson, Henrik},
  journal = {Phys. Rev. Lett.},
  volume = {118},
  issue = {1},
  pages = {015701},
  numpages = {5},
  year = {2017},
  month = {Jan},
  publisher = {American Physical Society},
  doi = {10.1103/PhysRevLett.118.015701},
  url = {https://link.aps.org/doi/10.1103/PhysRevLett.118.015701}
}

@article{PRB.2018.Sedlmayr,
  title = {Bulk-boundary correspondence for dynamical phase transitions in one-dimensional topological insulators and superconductors},
  author = {Sedlmayr, N. and Jaeger, P. and Maiti, M. and Sirker, J.},
  journal = {Phys. Rev. B},
  volume = {97},
  issue = {6},
  pages = {064304},
  numpages = {8},
  year = {2018},
  month = {Feb},
  publisher = {American Physical Society},
  doi = {10.1103/PhysRevB.97.064304},
  url = {https://link.aps.org/doi/10.1103/PhysRevB.97.064304}
}

@article{PRB.2019.Jafari,
  title = {Quench dynamics and zero-energy modes: The case of the Creutz model},
  author = {Jafari, R. and Johannesson, Henrik and Langari, A. and Martin-Delgado, M. A.},
  journal = {Phys. Rev. B},
  volume = {99},
  issue = {5},
  pages = {054302},
  numpages = {11},
  year = {2019},
  month = {Feb},
  publisher = {American Physical Society},
  doi = {10.1103/PhysRevB.99.054302},
  url = {https://link.aps.org/doi/10.1103/PhysRevB.99.054302}
}

@article{PRL.2019.Zache,
  title = {Dynamical Topological Transitions in the Massive Schwinger Model with a $\ensuremath{\theta}$ Term},
  author = {Zache, T. V. and Mueller, N. and Schneider, J. T. and Jendrzejewski, F. and Berges, J. and Hauke, P.},
  journal = {Phys. Rev. Lett.},
  volume = {122},
  issue = {5},
  pages = {050403},
  numpages = {6},
  year = {2019},
  month = {Feb},
  publisher = {American Physical Society},
  doi = {10.1103/PhysRevLett.122.050403},
  url = {https://link.aps.org/doi/10.1103/PhysRevLett.122.050403}
}

@article{PRB.2020.Masowski,
  title = {Quasiperiodic dynamical quantum phase transitions in multiband topological insulators and connections with entanglement entropy and fidelity susceptibility},
  author = {Mas\l{}owski, T. and Sedlmayr, N.},
  journal = {Phys. Rev. B},
  volume = {101},
  issue = {1},
  pages = {014301},
  numpages = {14},
  year = {2020},
  month = {Jan},
  publisher = {American Physical Society},
  doi = {10.1103/PhysRevB.101.014301},
  url = {https://link.aps.org/doi/10.1103/PhysRevB.101.014301}
}

@article{PRR.2021.Okugawa,
  title = {Mirror-symmetry-protected dynamical quantum phase transitions in topological crystalline insulators},
  author = {Okugawa, Ryo and Oshiyama, Hiroki and Ohzeki, Masayuki},
  journal = {Phys. Rev. Res.},
  volume = {3},
  issue = {4},
  pages = {043064},
  numpages = {9},
  year = {2021},
  month = {Oct},
  publisher = {American Physical Society},
  doi = {10.1103/PhysRevResearch.3.043064},
  url = {https://link.aps.org/doi/10.1103/PhysRevResearch.3.043064}
}

@article{PRA.2018.zhoulw,
  title = {Dynamical quantum phase transitions in non-Hermitian lattices},
  author = {Zhou, Longwen and Wang, Qing-hai and Wang, Hailong and Gong, Jiangbin},
  journal = {Phys. Rev. A},
  volume = {98},
  issue = {2},
  pages = {022129},
  numpages = {15},
  year = {2018},
  month = {Aug},
  publisher = {American Physical Society},
  doi = {10.1103/PhysRevA.98.022129},
  url = {https://link.aps.org/doi/10.1103/PhysRevA.98.022129}
}

@article{PRB.2023.Mondal,
  title = {Finite-temperature dynamical quantum phase transition in a non-Hermitian system},
  author = {Mondal, Debashish and Nag, Tanay},
  journal = {Phys. Rev. B},
  volume = {107},
  issue = {18},
  pages = {184311},
  numpages = {14},
  year = {2023},
  month = {May},
  publisher = {American Physical Society},
  doi = {10.1103/PhysRevB.107.184311},
  url = {https://link.aps.org/doi/10.1103/PhysRevB.107.184311}
}

@article{PRB.2022.Mondal,
  title = {Anomaly in the dynamical quantum phase transition in a non-Hermitian system with extended gapless phases},
  author = {Mondal, Debashish and Nag, Tanay},
  journal = {Phys. Rev. B},
  volume = {106},
  issue = {5},
  pages = {054308},
  numpages = {11},
  year = {2022},
  month = {Aug},
  publisher = {American Physical Society},
  doi = {10.1103/PhysRevB.106.054308},
  url = {https://link.aps.org/doi/10.1103/PhysRevB.106.054308}
}

@article{NJP.2021.zhoulongwen,
  title = {Non-{{Hermitian}} Topological Phases and Dynamical Quantum Phase Transitions: A Generic Connection},
  author = {Zhou, Longwen and Du, Qianqian},
  year = {2021},
  month = jun,
  journal = {New Journal of Physics},
  volume = {23},
  number = {6},
  pages = {063041},
  publisher = {IOP Publishing},
  doi = {10.1088/1367-2630/ac0574},
}

@article{EPL.2014.Sharma,
  title = {Loschmidt Echo and Dynamical Fidelity in Periodically Driven Quantum Systems},
  author = {Sharma, Shraddha and Russomanno, Angelo and Santoro, Giuseppe E. and Dutta, Amit},
  year = {2014},
  month = jun,
  journal = {Europhysics Letters},
  volume = {106},
  number = {6},
  pages = {67003},
  publisher = {{EDP Sciences, IOP Publishing and Societ{\`a} Italiana di Fisica}},
  doi = {10.1209/0295-5075/106/67003},
}

@article{PRB.2022.jafari,
  title = {Floquet dynamical quantum phase transitions under synchronized periodic driving},
  author = {Jafari, R. and Akbari, Alireza and Mishra, Utkarsh and Johannesson, Henrik},
  journal = {Phys. Rev. B},
  volume = {105},
  issue = {9},
  pages = {094311},
  numpages = {13},
  year = {2022},
  month = {Mar},
  publisher = {American Physical Society},
  doi = {10.1103/PhysRevB.105.094311},
  url = {https://link.aps.org/doi/10.1103/PhysRevB.105.094311}
}

@article{PRB.2019.YANGK,
  title = {Floquet dynamical quantum phase transitions},
  author = {Yang, Kai and Zhou, Longwen and Ma, Wenchao and Kong, Xi and Wang, Pengfei and Qin, Xi and Rong, Xing and Wang, Ya and Shi, Fazhan and Gong, Jiangbin and Du, Jiangfeng},
  journal = {Phys. Rev. B},
  volume = {100},
  issue = {8},
  pages = {085308},
  numpages = {11},
  year = {2019},
  month = {Aug},
  publisher = {American Physical Society},
  doi = {10.1103/PhysRevB.100.085308},
  url = {https://link.aps.org/doi/10.1103/PhysRevB.100.085308}
}

@article{PRB.2020.Zamani,
  title = {Floquet dynamical quantum phase transition in the extended XY model: Nonadiabatic to adiabatic topological transition},
  author = {Zamani, Sara and Jafari, R. and Langari, A.},
  journal = {Phys. Rev. B},
  volume = {102},
  issue = {14},
  pages = {144306},
  numpages = {12},
  year = {2020},
  month = {Oct},
  publisher = {American Physical Society},
  doi = {10.1103/PhysRevB.102.144306},
  url = {https://link.aps.org/doi/10.1103/PhysRevB.102.144306}
}

@article{JOP.2021.zhoulw,
  title = {Floquet Dynamical Quantum Phase Transitions in Periodically Quenched Systems},
  author = {Zhou, Longwen and Du, Qianqian},
  year = {2021},
  month = jul,
  journal = {Journal of Physics: Condensed Matter},
  volume = {33},
  number = {34},
  pages = {345403},
  publisher = {IOP Publishing},
  doi = {10.1088/1361-648X/ac0b60},
}

@article{PRA.2021.JAFARI,
  title = {Floquet dynamical phase transition and entanglement spectrum},
  author = {Jafari, R. and Akbari, Alireza},
  journal = {Phys. Rev. A},
  volume = {103},
  issue = {1},
  pages = {012204},
  numpages = {8},
  year = {2021},
  month = {Jan},
  publisher = {American Physical Society},
  doi = {10.1103/PhysRevA.103.012204},
  url = {https://link.aps.org/doi/10.1103/PhysRevA.103.012204}
}

@article{NatureC.2021.Hamazaki,
  title = {Exceptional Dynamical Quantum Phase Transitions in Periodically Driven Systems},
  author = {Hamazaki, Ryusuke},
  year = {2021},
  month = sep,
  journal = {Nature Communications},
  volume = {12},
  number = {1},
  pages = {5108},
  issn = {2041-1723},
  doi = {10.1038/s41467-021-25355-3},
}

@article{PRB.2022.Zamani,
  title = {Out-of-time-order correlations and Floquet dynamical quantum phase transition},
  author = {Zamani, Sara and Jafari, R. and Langari, A.},
  journal = {Phys. Rev. B},
  volume = {105},
  issue = {9},
  pages = {094304},
  numpages = {15},
  year = {2022},
  month = {Mar},
  publisher = {American Physical Society},
  doi = {10.1103/PhysRevB.105.094304},
  url = {https://link.aps.org/doi/10.1103/PhysRevB.105.094304}
}

@article{PhysicaA.2022.Luan,
  title = {Floquet Dynamical Quantum Phase Transitions of the {{XY}} Spin-Chain under Periodic Quenching},
  author = {Luan, Li-Na and Zhang, Mei-Yu and Wang, L.C.},
  year = {2022},
  journal = {Physica A: Statistical Mechanics and its Applications},
  volume = {604},
  pages = {127866},
  issn = {0378-4371},
  doi = {10.1016/j.physa.2022.127866},
}

@article{PRB.2018.Lang,
  title = {Dynamical quantum phase transition for mixed states in open systems},
  author = {Lang, Haifeng and Chen, Yixin and Hong, Qiantan and Fan, Heng},
  journal = {Phys. Rev. B},
  volume = {98},
  issue = {13},
  pages = {134310},
  numpages = {7},
  year = {2018},
  month = {Oct},
  publisher = {American Physical Society},
  doi = {10.1103/PhysRevB.98.134310},
  url = {https://link.aps.org/doi/10.1103/PhysRevB.98.134310}
}

@article{ScientificReports.2018.Bandyopadhyay,
  title = {Exploring the Possibilities of Dynamical Quantum Phase Transitions in the Presence of a {{Markovian}} Bath},
  author = {Bandyopadhyay, Souvik and Laha, Sudarshana and Bhattacharya, Utso and Dutta, Amit},
  year = {2018},
  month = aug,
  journal = {Scientific Reports},
  volume = {8},
  number = {1},
  pages = {11921},
  issn = {2045-2322},
  doi = {10.1038/s41598-018-30377-x},
}

@article{PRB.2020.Hou,
  title = {Ubiquity of zeros of the Loschmidt amplitude for mixed states in different physical processes and its implication},
  author = {Hou, Xu-Yang and Gao, Qu-Cheng and Guo, Hao and He, Yan and Liu, Tong and Chien, Chih-Chun},
  journal = {Phys. Rev. B},
  volume = {102},
  issue = {10},
  pages = {104305},
  numpages = {13},
  year = {2020},
  month = {Sep},
  publisher = {American Physical Society},
  doi = {10.1103/PhysRevB.102.104305},
  url = {https://link.aps.org/doi/10.1103/PhysRevB.102.104305}
}

@article{PRA.2020.Kyaw,
  title = {Dynamical quantum phase transitions and non-Markovian dynamics},
  author = {Kyaw, Thi Ha and Bastidas, Victor M. and Tangpanitanon, Jirawat and Romero, Guillermo and Kwek, Leong-Chuan},
  journal = {Phys. Rev. A},
  volume = {101},
  issue = {1},
  pages = {012111},
  numpages = {10},
  year = {2020},
  month = {Jan},
  publisher = {American Physical Society},
  doi = {10.1103/PhysRevA.101.012111},
  url = {https://link.aps.org/doi/10.1103/PhysRevA.101.012111}
}

@article{PRB.2018.Mera,
  title = {Dynamical phase transitions at finite temperature from fidelity and interferometric Loschmidt echo induced metrics},
  author = {Mera, Bruno and Vlachou, Chrysoula and Paunkovi\ifmmode \acute{c}\else \'{c}\fi{}, Nikola and Vieira, V\'{\i}tor R. and Viyuela, Oscar},
  journal = {Phys. Rev. B},
  volume = {97},
  issue = {9},
  pages = {094110},
  numpages = {15},
  year = {2018},
  month = {Mar},
  publisher = {American Physical Society},
  doi = {10.1103/PhysRevB.97.094110},
  url = {https://link.aps.org/doi/10.1103/PhysRevB.97.094110}
}

@article{PRB.2018.SedlmayrFate,
  title = {Fate of dynamical phase transitions at finite temperatures and in open systems},
  author = {Sedlmayr, N. and Fleischhauer, M. and Sirker, J.},
  journal = {Phys. Rev. B},
  volume = {97},
  issue = {4},
  pages = {045147},
  numpages = {8},
  year = {2018},
  month = {Jan},
  publisher = {American Physical Society},
  doi = {10.1103/PhysRevB.97.045147},
  url = {https://link.aps.org/doi/10.1103/PhysRevB.97.045147}
}

@article{PRL.2020.Link,
  title = {Dynamical Phase Transitions in Dissipative Quantum Dynamics with Quantum Optical Realization},
  author = {Link, Valentin and Strunz, Walter T.},
  journal = {Phys. Rev. Lett.},
  volume = {125},
  issue = {14},
  pages = {143602},
  numpages = {6},
  year = {2020},
  month = {Sep},
  publisher = {American Physical Society},
  doi = {10.1103/PhysRevLett.125.143602},
  url = {https://link.aps.org/doi/10.1103/PhysRevLett.125.143602}
}

@article{PRA.2021.Hou,
  title = {Finite-temperature topological phase transitions of spin-$j$ systems in Uhlmann processes: General formalism and experimental protocols},
  author = {Hou, Xu-Yang and Guo, Hao and Chien, Chih-Chun},
  journal = {Phys. Rev. A},
  volume = {104},
  issue = {2},
  pages = {023303},
  numpages = {14},
  year = {2021},
  month = {Aug},
  publisher = {American Physical Society},
  doi = {10.1103/PhysRevA.104.023303},
  url = {https://link.aps.org/doi/10.1103/PhysRevA.104.023303}
}

@article{PRB.2013.Karrasch,
  title = {Dynamical phase transitions after quenches in nonintegrable models},
  author = {Karrasch, C. and Schuricht, D.},
  journal = {Phys. Rev. B},
  volume = {87},
  issue = {19},
  pages = {195104},
  numpages = {8},
  year = {2013},
  month = {May},
  publisher = {American Physical Society},
  doi = {10.1103/PhysRevB.87.195104},
  url = {https://link.aps.org/doi/10.1103/PhysRevB.87.195104}
}

@article{PRB.2014.Andraschko,
  title = {Dynamical quantum phase transitions and the Loschmidt echo: A transfer matrix approach},
  author = {Andraschko, F. and Sirker, J.},
  journal = {Phys. Rev. B},
  volume = {89},
  issue = {12},
  pages = {125120},
  numpages = {13},
  year = {2014},
  month = {Mar},
  publisher = {American Physical Society},
  doi = {10.1103/PhysRevB.89.125120},
  url = {https://link.aps.org/doi/10.1103/PhysRevB.89.125120}
}

@article{PRB.2014.Kriel,
  title = {Dynamical quantum phase transitions in the axial next-nearest-neighbor Ising chain},
  author = {Kriel, J. N. and Karrasch, C. and Kehrein, S.},
  journal = {Phys. Rev. B},
  volume = {90},
  issue = {12},
  pages = {125106},
  numpages = {9},
  year = {2014},
  month = {Sep},
  publisher = {American Physical Society},
  doi = {10.1103/PhysRevB.90.125106},
  url = {https://link.aps.org/doi/10.1103/PhysRevB.90.125106}
}

@article{PRB.2015.Sharma,
  title = {Quenches and dynamical phase transitions in a nonintegrable quantum Ising model},
  author = {Sharma, Shraddha and Suzuki, Sei and Dutta, Amit},
  journal = {Phys. Rev. B},
  volume = {92},
  issue = {10},
  pages = {104306},
  numpages = {7},
  year = {2015},
  month = {Sep},
  publisher = {American Physical Society},
  doi = {10.1103/PhysRevB.92.104306},
  url = {https://link.aps.org/doi/10.1103/PhysRevB.92.104306}
}

@article{PRB.2017.Homrighausen,
  title = {Anomalous dynamical phase in quantum spin chains with long-range interactions},
  author = {Homrighausen, Ingo and Abeling, Nils O. and Zauner-Stauber, Valentin and Halimeh, Jad C.},
  journal = {Phys. Rev. B},
  volume = {96},
  issue = {10},
  pages = {104436},
  numpages = {7},
  year = {2017},
  month = {Sep},
  publisher = {American Physical Society},
  doi = {10.1103/PhysRevB.96.104436},
  url = {https://link.aps.org/doi/10.1103/PhysRevB.96.104436}
}

@article{PRB.2017.Obuchi,
  title = {Complex semiclassical analysis of the Loschmidt amplitude and dynamical quantum phase transitions},
  author = {Obuchi, Tomoyuki and Suzuki, Sei and Takahashi, Kazutaka},
  journal = {Phys. Rev. B},
  volume = {95},
  issue = {17},
  pages = {174305},
  numpages = {11},
  year = {2017},
  month = {May},
  publisher = {American Physical Society},
  doi = {10.1103/PhysRevB.95.174305},
  url = {https://link.aps.org/doi/10.1103/PhysRevB.95.174305}
}

@article{PRB.2017.Dutta,
  title = {Probing the role of long-range interactions in the dynamics of a long-range Kitaev chain},
  author = {Dutta, Anirban and Dutta, Amit},
  journal = {Phys. Rev. B},
  volume = {96},
  issue = {12},
  pages = {125113},
  numpages = {8},
  year = {2017},
  month = {Sep},
  publisher = {American Physical Society},
  doi = {10.1103/PhysRevB.96.125113},
  url = {https://link.aps.org/doi/10.1103/PhysRevB.96.125113}
}

@article{PRR.2020.Halimeh,
  title = {Quasiparticle origin of dynamical quantum phase transitions},
  author = {Halimeh, Jad C. and Van Damme, Maarten and Zauner-Stauber, Valentin and Vanderstraeten, Laurens},
  journal = {Phys. Rev. Res.},
  volume = {2},
  issue = {3},
  pages = {033111},
  numpages = {8},
  year = {2020},
  month = {Jul},
  publisher = {American Physical Society},
  doi = {10.1103/PhysRevResearch.2.033111},
  url = {https://link.aps.org/doi/10.1103/PhysRevResearch.2.033111}
}

@article{PRB.2017.Weidinger,
  title = {Dynamical quantum phase transitions in systems with continuous symmetry breaking},
  author = {Weidinger, Simon A. and Heyl, Markus and Silva, Alessandro and Knap, Michael},
  journal = {Phys. Rev. B},
  volume = {96},
  issue = {13},
  pages = {134313},
  numpages = {8},
  year = {2017},
  month = {Oct},
  publisher = {American Physical Society},
  doi = {10.1103/PhysRevB.96.134313},
  url = {https://link.aps.org/doi/10.1103/PhysRevB.96.134313}
}

@article{PRL.2017.Jurcevic,
  title = {Direct Observation of Dynamical Quantum Phase Transitions in an Interacting Many-Body System},
  author = {Jurcevic, P. and Shen, H. and Hauke, P. and Maier, C. and Brydges, T. and Hempel, C. and Lanyon, B. P. and Heyl, M. and Blatt, R. and Roos, C. F.},
  journal = {Phys. Rev. Lett.},
  volume = {119},
  issue = {8},
  pages = {080501},
  numpages = {5},
  year = {2017},
  month = {Aug},
  publisher = {American Physical Society},
  doi = {10.1103/PhysRevLett.119.080501},
  url = {https://link.aps.org/doi/10.1103/PhysRevLett.119.080501}
}

@article{PRAp.2019.GUOXY,
  title = {Observation of a Dynamical Quantum Phase Transition by a Superconducting Qubit Simulation},
  author = {Guo, Xue-Yi and Yang, Chao and Zeng, Yu and Peng, Yi and Li, He-Kang and Deng, Hui and Jin, Yi-Rong and Chen, Shu and Zheng, Dongning and Fan, Heng},
  journal = {Phys. Rev. Appl.},
  volume = {11},
  issue = {4},
  pages = {044080},
  numpages = {12},
  year = {2019},
  month = {Apr},
  publisher = {American Physical Society},
  doi = {10.1103/PhysRevApplied.11.044080},
  url = {https://link.aps.org/doi/10.1103/PhysRevApplied.11.044080}
}

@article{PRL.2019.Wang,
  title = {Simulating Dynamic Quantum Phase Transitions in Photonic Quantum Walks},
  author = {Wang, Kunkun and Qiu, Xingze and Xiao, Lei and Zhan, Xiang and Bian, Zhihao and Yi, Wei and Xue, Peng},
  journal = {Phys. Rev. Lett.},
  volume = {122},
  issue = {2},
  pages = {020501},
  numpages = {6},
  year = {2019},
  month = {Jan},
  publisher = {American Physical Society},
  doi = {10.1103/PhysRevLett.122.020501},
  url = {https://link.aps.org/doi/10.1103/PhysRevLett.122.020501}
}

@article{Light.2020.Xu,
  title = {Measuring a Dynamical Topological Order Parameter in Quantum Walks},
  author = {Xu, Xiao-Ye and Wang, Qin-Qin and Heyl, Markus and Budich, Jan Carl and Pan, Wei-Wei and Chen, Zhe and Jan, Munsif and Sun, Kai and Xu, Jin-Shi and Han, Yong-Jian and Li, Chuan-Feng and Guo, Guang-Can},
  year = 2020,
  month = jan,
  journal = {Light: Science \& Applications},
  volume = {9},
  number = {1},
  pages = {7},
  issn = {2047-7538},
  doi = {10.1038/s41377-019-0237-8},
}

@article{LSC.2025.Zhang,
  title = {Self-Normal and Biorthogonal Dynamical Quantum Phase Transitions in Non-{{Hermitian}} Quantum Walks},
  author = {Zhang, Haiting and Wang, Kunkun and Xiao, Lei and Xue, Peng},
  year = 2025,
  month = jul,
  journal = {Light: Science \& Applications},
  volume = {14},
  number = {1},
  pages = {253},
  issn = {2047-7538},
  doi = {10.1038/s41377-025-01919-6},
}

@article{PRB.2019.Abdi,
  title = {Dynamical quantum phase transition in Bose-Einstein condensates},
  author = {Abdi, Mehdi},
  journal = {Phys. Rev. B},
  volume = {100},
  issue = {18},
  pages = {184310},
  numpages = {10},
  year = {2019},
  month = {Nov},
  publisher = {American Physical Society},
  doi = {10.1103/PhysRevB.100.184310},
  url = {https://link.aps.org/doi/10.1103/PhysRevB.100.184310}
}

@article{APL.2020.Chen,
  title = {Detecting the Out-of-Time-Order Correlations of Dynamical Quantum Phase Transitions in a Solid-State Quantum Simulator},
  author = {Chen, Bing and Hou, Xianfei and Zhou, Feifei and Qian, Peng and Shen, Heng and Xu, Nanyang},
  year = 2020,
  month = may,
  journal = {Applied Physics Letters},
  volume = {116},
  number = {19},
  pages = {194002},
  issn = {0003-6951, 1077-3118},
  doi = {10.1063/5.0004152},
  urldate = {2025-10-20},
}

@article{NaturePhysics.2018.flaschner,
  title = {Observation of Dynamical Vortices after Quenches in a System with Topology},
  author = {Fl{\"a}schner, N. and Vogel, D. and Tarnowski, M. and Rem, B. S. and L{\"u}hmann, D.-S. and Heyl, M. and Budich, J. C. and Mathey, L. and Sengstock, K. and Weitenberg, C.},
  year = {2018},
  month = mar,
  journal = {Nature Physics},
  volume = {14},
  number = {3},
  pages = {265--268},
  issn = {1745-2481},
  doi = {10.1038/s41567-017-0013-8}}

@article{PRA.2018.XU,
  title = {Effects of quantum coherence on work statistics},
  author = {Xu, Bao-Ming and Zou, Jian and Guo, Li-Sha and Kong, Xiang-Mu},
  journal = {Phys. Rev. A},
  volume = {97},
  issue = {5},
  pages = {052122},
  numpages = {12},
  year = {2018},
  month = {May},
  publisher = {American Physical Society},
  doi = {10.1103/PhysRevA.97.052122},
  url = {https://link.aps.org/doi/10.1103/PhysRevA.97.052122}
}

@article{CTP.2024.XU,
  title = {Quantum Coherence Assisted Dynamical Phase Transition},
  author = {Xu, Bao-Ming},
  year = {2024},
  month = dec,
  journal = {Communications in Theoretical Physics},
  volume = {76},
  number = {12},
  pages = {125104},
  issn = {0253-6102, 1572-9494},
  doi = {10.1088/1572-9494/ad724c},
  urldate = {2025-09-20},
}

@article{Annual.2018.Mitra,
   title={Quantum Quench Dynamics},
   volume={9},
   ISSN={1947-5462},
   url={http://dx.doi.org/10.1146/annurev-conmatphys-031016-025451},
   DOI={10.1146/annurev-conmatphys-031016-025451},
   number={1},
   journal={Annual Review of Condensed Matter Physics},
   publisher={Annual Reviews},
   author={Mitra, Aditi},
   year={2018},
   month=mar, pages={245–259} }

@article{2016Essler,
doi = {10.1088/1742-5468/2016/06/064002},
url = {https://doi.org/10.1088/1742-5468/2016/06/064002},
year = {2016},
month = {jun},
publisher = {IOP Publishing and SISSA},
volume = {2016},
number = {6},
pages = {064002},
author = {Essler, Fabian H L and Fagotti, Maurizio},
title = {Quench dynamics and relaxation in isolated integrable quantum spin chains},
journal = {Journal of Statistical Mechanics: Theory and Experiment}
}

@article{PRB.2020.Uhrich,
  title = {Out-of-equilibrium phase diagram of long-range superconductors},
  author = {Uhrich, Philipp and Defenu, Nicol\`o and Jafari, Rouhollah and Halimeh, Jad C.},
  journal = {Phys. Rev. B},
  volume = {101},
  issue = {24},
  pages = {245148},
  numpages = {15},
  year = {2020},
  month = {Jun},
  publisher = {American Physical Society},
  doi = {10.1103/PhysRevB.101.245148},
  url = {https://link.aps.org/doi/10.1103/PhysRevB.101.245148}
}

@article{PRR.2022.Hashizume,
  title = {Dynamical phase transitions in the two-dimensional transverse-field Ising model},
  author = {Hashizume, Tomohiro and McCulloch, Ian P. and Halimeh, Jad C.},
  journal = {Phys. Rev. Res.},
  volume = {4},
  issue = {1},
  pages = {013250},
  numpages = {9},
  year = {2022},
  month = {Mar},
  publisher = {American Physical Society},
  doi = {10.1103/PhysRevResearch.4.013250},
  url = {https://link.aps.org/doi/10.1103/PhysRevResearch.4.013250}
}

@article{PRB.2018.Lang.qcp,
  title = {Concurrence of dynamical phase transitions at finite temperature in the fully connected transverse-field Ising model},
  author = {Lang, Johannes and Frank, Bernhard and Halimeh, Jad C.},
  journal = {Phys. Rev. B},
  volume = {97},
  issue = {17},
  pages = {174401},
  numpages = {15},
  year = {2018},
  month = {May},
  publisher = {American Physical Society},
  doi = {10.1103/PhysRevB.97.174401},
  url = {https://link.aps.org/doi/10.1103/PhysRevB.97.174401}
}

@article{PRB.2022.Rossi,
  title = {Nonlinear current and dynamical quantum phase transitions in the flux-quenched Su-Schrieffer-Heeger model},
  author = {Rossi, Lorenzo and Dolcini, Fabrizio},
  journal = {Phys. Rev. B},
  volume = {106},
  issue = {4},
  pages = {045410},
  numpages = {8},
  year = {2022},
  month = {Jul},
  publisher = {American Physical Society},
  doi = {10.1103/PhysRevB.106.045410},
  url = {https://link.aps.org/doi/10.1103/PhysRevB.106.045410}
}

@Article{SciPostPhys.2021.Liska,
	title={{The Loschmidt Index}},
	author={Diego Liska and Vladimir Gritsev},
	journal={SciPost Phys.},
	volume={10},
	pages={100},
	year={2021},
	publisher={SciPost},
	doi={10.21468/SciPostPhys.10.5.100},
	url={https://scipost.org/10.21468/SciPostPhys.10.5.100},
}

@article{PRB.2024.Wong,
  title = {Entanglement in quenched extended Su-Schrieffer-Heeger model with anomalous dynamical quantum phase transitions},
  author = {Wong, Cheuk Yiu and Hui, Tsz Hin and Sacramento, P. D. and Yu, Wing Chi},
  journal = {Phys. Rev. B},
  volume = {110},
  issue = {5},
  pages = {054312},
  numpages = {17},
  year = {2024},
  month = {Aug},
  publisher = {American Physical Society},
  doi = {10.1103/PhysRevB.110.054312},
  url = {https://link.aps.org/doi/10.1103/PhysRevB.110.054312}
}

@article{PRL.2016.Huang,
  title = {Dynamical Quantum Phase Transitions: Role of Topological Nodes in Wave Function Overlaps},
  author = {Huang, Zhoushen and Balatsky, Alexander V.},
  journal = {Phys. Rev. Lett.},
  volume = {117},
  issue = {8},
  pages = {086802},
  numpages = {6},
  year = {2016},
  month = {Aug},
  publisher = {American Physical Society},
  doi = {10.1103/PhysRevLett.117.086802},
  url = {https://link.aps.org/doi/10.1103/PhysRevLett.117.086802}
}

@article{PRB.2018.Bhattacharjee,
  title = {Dynamical quantum phase transitions in extended transverse Ising models},
  author = {Bhattacharjee, Sourav and Dutta, Amit},
  journal = {Phys. Rev. B},
  volume = {97},
  issue = {13},
  pages = {134306},
  numpages = {8},
  year = {2018},
  month = {Apr},
  publisher = {American Physical Society},
  doi = {10.1103/PhysRevB.97.134306},
  url = {https://link.aps.org/doi/10.1103/PhysRevB.97.134306}
}

@article{PRB.2019.Zhou,
  title = {Signature of a nonequilibrium quantum phase transition in the long-time average of the Loschmidt echo},
  author = {Zhou, Bozhen and Yang, Chao and Chen, Shu},
  journal = {Phys. Rev. B},
  volume = {100},
  issue = {18},
  pages = {184313},
  numpages = {7},
  year = {2019},
  month = {Nov},
  publisher = {American Physical Society},
  doi = {10.1103/PhysRevB.100.184313},
  url = {https://link.aps.org/doi/10.1103/PhysRevB.100.184313}
}

@article{PRL.2023.Corps,
  title = {Theory of Dynamical Phase Transitions in Quantum Systems with Symmetry-Breaking Eigenstates},
  author = {Corps, \'Angel L. and Rela\~no, Armando},
  journal = {Phys. Rev. Lett.},
  volume = {130},
  issue = {10},
  pages = {100402},
  numpages = {7},
  year = {2023},
  month = {Mar},
  publisher = {American Physical Society},
  doi = {10.1103/PhysRevLett.130.100402},
  url = {https://link.aps.org/doi/10.1103/PhysRevLett.130.100402}
}

@article{PRB.2014.Hickey,
  title = {Dynamical phase transitions, time-integrated observables, and geometry of states},
  author = {Hickey, James M. and Genway, Sam and Garrahan, Juan P.},
  journal = {Phys. Rev. B},
  volume = {89},
  issue = {5},
  pages = {054301},
  numpages = {9},
  year = {2014},
  month = {Feb},
  publisher = {American Physical Society},
  doi = {10.1103/PhysRevB.89.054301},
  url = {https://link.aps.org/doi/10.1103/PhysRevB.89.054301}
}

@article{PFEUTY197079,
  title = {The One-Dimensional {{Ising}} Model with a Transverse Field},
  author = {Pfeuty, Pierre},
  year = 1970,
  journal = {Annals of Physics},
  volume = {57},
  number = {1},
  pages = {79--90},
  issn = {0003-4916},
  doi = {10.1016/0003-4916(70)90270-8},
}
\end{document}